\documentclass[ %
 reprint,
 amsmath,amssymb,
 aps,
 prd,
 floatfix,
]{revtex4-2}

\usepackage{mathrsfs}
\usepackage{graphicx}
\usepackage{dcolumn}
\usepackage{bm}
\usepackage[colorlinks=true, allcolors=blue]{hyperref}

\hypersetup{
    pdftitle={Dendro-GR at high mass ratios with high spins},
    pdfauthor={William K. Black et al.}, 
    pdfsubject={BBH}, 
    pdfstartview={FitH}, 
    pdfkeywords={Numerical Relativity, Gravitational waves, Gravitational wave sources, Classical black holes, General relativity, Dendro, Dendro-GR, supercomputing, black holes, binary inspirals, BBH, gravitational waves, WAMR, ORBIT, high mass ratio, high spin}
}

\usepackage{soul, float}

\usepackage[dvipsnames]{xcolor}

\definecolor{bleudefrance}{rgb}{0.19, 0.55, 0.91}
\definecolor{AmericanRed}{rgb}{0.698, 0.133, 0.204}
\definecolor{AmericanBlue}{rgb}{0.0391, 0.1914, 0.3789}
\definecolor{mustard}{rgb}{.808,.702,.00392}
\definecolor{ketchup}{rgb}{0.781, 0.160, 0.1328}
\definecolor{beeswax}{rgb}{.9137, .6235, .2471} 
\definecolor{PumpkinOrange}{rgb}{0.984, 0.49, 0.027}

\newcommand{\BSSN}{BSSN}
\newcommand{\dendro}{\textsc{Dendro}}
\newcommand{\dendrogr}{{\textsc{Dendro-GR}}}

\newcommand{\bfx}{{\bf x}}

\begin{document}


\title{
  {\dendrogr} at high mass ratios with high spins
}

\author{\href{https://orcid.org/0000-0003-4811-7913}{William K. Black}}
  \email{wblack@uidaho.edu}
\author{\href{https://orcid.org/0000-0002-6142-5542}{David Neilsen}}
  \email{david.neilsen@byu.edu}
\author{\href{https://orcid.org/0000-0003-2164-8001}{Eric W. Hirschmann}}
  \email{ehirsch@byu.edu}
\affiliation{
 Department of Physics and Astronomy, 
 Brigham Young University, \\
 N283 Carl F. Eyring Science Center, Provo, UT 84602
}

\author{\href{https://orcid.org/0000-0003-0610-0806}{David F. Van Komen}}
  \email{david.vankomen@gmail.com}
\affiliation{
  Kahlert School of Computing, 
  University of Utah, 
  50 Central Campus Dr., Room 3190, Salt Lake City, UT 84112
}

\date{\today}

\begin{abstract}

The Laser Interferometer Space Antenna (LISA) launches in less than a decade; 
it will detect spinning high-mass-ratio binary black hole inspirals annually, 
alongside other third-generation gravitational wave detectors. 
High-mass-ratio systems occupy a regime where numerical-relativity simulations 
remain computationally expensive and technically demanding, 
especially with high spins at precessing orientations. 
This portion of parameter space thus remains undersampled,
leading to significant bias in parameter estimation. 
We must close these gaps. 

Here we report key progress in \dendrogr\ toward reducing the computational
cost of high-mass-ratio binaries with spin.
We evolve the first \dendrogr\ binaries at mass ratio $q=24$ (nonspinning) 
and at $q=12$ with spins up to $\chi=0.8$ on both black holes, spanning various configurations. 
These proof-of-concept runs show
strong evidence that \dendrogr\ can simulate in this regime and beyond. 
The simulations resolve the expected multipolar hierarchy in the gravitational waveforms 
through multipole modes up to $\ell=8$, remain stable, keep constraint violations low 
and largely constant, conserve horizon mass, and 
have high computational efficiency with relatively low wall-hour cost. 
These results establish our starting line for systematic exploration of 
the high-mass-ratio, high-spin binary black hole systems 
that are needed for gravitational wave analysis.
\end{abstract}


\maketitle


\section{Introduction} \label{sec:intro}

The advent and development of gravitational wave astronomy over the last decade has been nothing short of remarkable, 
providing heretofore inaccessible insights and understanding of the high-energy universe~\citep[see e.g.][]{LIGOScientific:2017vwq, LIGOScientific:2025rid}.
With published detections from the LIGO-Virgo-KAGRA (LVK) collaboration approaching 400, we continue to grow our collection of known binary mergers and their properties~\citep{theligoscientificcollaboration2026gwtc50introductionversion50}.  
After these spectacular successes, next generation detectors have been proposed,
such as the Einstein Telescope~\citep{Maggiore_2020} and Cosmic Explorer~\citep{reitze2019cosmicexploreruscontribution}.  
The LISA (Laser Interferometer Space Antenna) mission, in particular, 
is in active development with an expected launch date of the mid-2030s \citep{amaroseoane2017laserinterferometerspaceantenna}.  

LISA and third-generation (3G) ground-based detectors will open new frequency bands to gravitational wave astronomy and provide much higher sensitivities for a variety of compact object mergers~\cite{Gupta:2023lga}.  
This increase in detection sensitivity will necessarily require commensurate improvements in precision for source modeling and the simulation of waveforms~\citep{Purrer_Haster_20, Ferguson+21, Jan_Ferguson+24, Thompson:2025hhc}. 
Numerical relativity simulations in currently published catalogs 
have, on average, sufficiently low systematic errors (simulation bias) for typical LVK data analysis. 
However, high-precision tests of general relativity
in the strong-field regime will require a reduction of these errors~\cite{Mezzasoma+26}.
Moreover, current catalogs are of insufficient quality for LISA and other future 3G detectors,
and it is broadly recognized that this could significantly bias the interpretation and parameter estimation of waveforms, potentially
limiting the effectiveness of 3G detectors unless higher accuracy simulations are achieved. 

Perhaps not as well appreciated is the fact that the current library of waveforms 
insufficiently samples certain portions of the binary black hole (BBH) parameter space. 
Recent work of~\citet{Mahapatra+26} shows that current waveform models used in parameter estimation can substantially mischaracterize a strongly precessing, highly spinning ($\chi_1 = 0.8$), large-mass-ratio ($q=18$, where $q \equiv m_1/m_2 \ge 1$) binary, erroneously recovering a mass ratio below $q=10$, underestimating the primary spin, with some cases showing over 100\% error in the estimated mass of the secondary. 
\citet{Pompili+23} similarly shows significant mismatch between models and simulations at high mass ratios, particularly in systems with substantial spin. 
To properly recognize observed signals from high-mass-ratio inspirals reaching 3G detectors, 
we must fill this portion of parameter space---high-mass-ratio systems with high spin---including strongly precessing cases.  

To date, relatively few simulations exist at mass ratios $q \gtrsim 10$. 
\citet{Gonzalez+09} evolved the first full $q=10$ inspiral using the BAM code 
(recently used for the $q=18$ suite described above). 
Since then, several additional groups have pushed beyond $q=10$. 
Using the SpEC code, the SXS collaboration has simulated over 90 publicly available configurations with $10 < q \leq 15$ and one with $q=20$.  
All have at least 20 orbits prior to merger. 
None have spin on the secondary BH, though many  
have precessing spins~\citep{SXSPackage_v2025.0.19, SXSCatalogPaper_3, SXSCatalogData_3.0.0}. 
The RIT group, using the LazEv code built on 
the {\sc Cactus} / {\sc Carpet} / {\sc Einstein Toolkit} infrastructure, 
has simulated 36 quasicircular inspirals for mass ratios of $q > 10$, 
publishing waveforms up to $q=128$~\citep{Healy_Lousto_2022}. 
The median inspiral has 13 orbits to merger; 
several have significant spin magnitudes. 
The MAYA catalog has ten runs at $q>10$, 
with a median of $\sim 5$ orbits at a variety of spin orientations~\citep{Ferguson+25}. 
While around ten thousand comparable mass ratio ($q<10$) simulations have been published, 
large mass ratios ($q>10$) have fewer than 200 published waveforms, and
often for relatively few orbits (which can limit their utility).
Many current codes struggle to access this regime of parameter space;
as a result, the high-$q$ parameter space remains quite sparsely populated.  

The paucity of large-mass-ratio simulations is driven by their higher computational costs. 
Larger mass ratios introduce progressively smaller length scales to the problem---the horizon size of the secondary black hole---demanding additional refinement and correspondingly smaller time steps.  
The complete computational cost is compounded by other factors, 
and taken together they quickly threaten to become prohibitive as $q$ increases~\cite{LISAConsortiumWaveformWorkingGroup:2023arg}.
For example, the spatial resolution $\Delta x$ and time step $\Delta t$ both scale as the mass of the secondary $m_2 = 1/(1 + q)$;
so the number of steps per unit time scales as $1/\Delta t \sim q$. 
Post-Newtonian estimates show that the time-to-merger of binary systems scales as 
$(1+q)^2/q$~\cite{Maggiore_08}.  
The number of points on the grid scales as  
$(1/\Delta x)^3 \sim q^3$ on a uniform grid, 
but an idealized adaptive mesh refinement scheme
can reduce this to $\mathcal{O}(\log_2 q)$ \footnote{
  Assuming self-similar structure, each halving of the smaller BH's mass leads to an identical increase in mesh size (octree leaves on the grid, proportional to total point count). Because the mass of the smaller BH $m$ relates to mass ratio $q$ as $m = 1/(1 + q)$, we expect point count to roughly increase with mass ratio as $\log_2 (1 + q)$, which approaches $\log_2 q$ for $q \gg 1$. 
}. 
Taken together, the overall cost of a $q\gg 1$ simulation will effectively scale with mass ratio as $\sim q^2 \log_2 q$ (assuming global time stepping and a fixed initial separation $D_0/M$).  
Longer runs may require even more refinement to control the accumulated phase error
in the waveform during the inspiral and merger. 
Black hole spin reduces the horizon size and warps nearby spacetime, 
requiring additional refinement toward large spins.
Clearly, simulating high-$q$ BBH systems and their resulting waveforms to sufficient accuracy is quite challenging and there are few simple solutions.  
Such challenging simulations will require several innovative strategies to enable them.

The scaling argument above shows that highly efficient
adaptive refinement is essential to enable high-$q$ numerical relativity simulations, 
yet there remains some uncertainty on the best approaches in employing it---how to 
determine refinement criteria~\citep{Radia+22, Rashti+24a}. 
The gravitational waveform is usually the most important result of a particular
simulation, yet it is difficult to correlate errors in the waveform to specific
refinement choices made during the simulation. The waveform, after all,
is extracted far from the dynamical regions of the grid and well after
the times when refinement choices must be made.  

This work presents some of our efforts towards addressing these issues.
Using \href{https://github.com/paralab/Dendro-GR/}{\dendrogr}, we show that refinement criteria related to the local convergence 
of the solution are important for reducing waveform noise.
Moreover, we show that the regions requiring refinement may at times be non-intuitive; 
this provides evidence that unstructured grids are more efficient than nested refinement schemes.
{\dendrogr} is built on the parallel octree structure of \dendro, 
extended to hyperbolic PDEs.  {\dendro} uses unstructured grids with octrees as the primary data structure~\cite{Sundar+08}, with octree partitioning based on space-filling curves~\cite{Fernando+17} to ensure scalability~\cite{Fernando+19b}.  
For refinement, we use Wavelet Adaptive Multiresolution Representation (WAMR)~\cite{Paolucci1, Paolucci2, DeBuhr+18, Fernando+23}; 
this basis of iterated, interpolating wavelets creates a sparse, unstructured grid that naturally adapts to the dynamical features of the solution. 
The coefficients in the wavelet expansion provide 
a direct measure of the local error of the expansion, 
thereby providing an automatic refinement
criterion~\cite{Bertoluzza1996, Vasilyev2000, Regele2009, Vasilyev1995, Vasilyev1996, Vasilyev1997}. 
For the current work, \dendrogr\ solves the vacuum Einstein equations using the conventional second-order \BSSN~\cite{1987PThPS..90....1N, 1995PhRvD..52.5428S, 1999PhRvD..59b4007B, lousto} formulation with standard moving puncture gauge as its base.  
\dendrogr\ has demonstrated excellent scaling up to and beyond $10^5$ processors~\cite{Fernando+19b}, 
enabling efficient runs of large mass ratio BBHs.  
The \dendrogr\ collaboration has published inspiral waveforms for spinless mass ratios $4 \leq q \le 16$~\cite{Fernando+23}.

More recently,
we have tuned \dendrogr 's settings for compatibility with increasingly large mass ratios and spins. 
The current work showcases \dendrogr's continued development with several proof-of-concept runs, 
as we move simultaneously toward 
increasingly large mass ratios (up to $q=24$)
and large spins (up to $\chi = 0.8$),
including a strongly precessing case. 
In \S\ref{sec:methods}, we introduce some of our code changes which have permitted progress towards
higher $q$ and $\chi$.
We then present waveforms and cross-catalog comparisons in \S\ref{sec:results},
and in \S\ref{sec:discuss_WAMR} we 
discuss the benefits of WAMR (feature-based refinement). 
To the extent that we have experimented, these simulations do not appear to be the bounds on \dendrogr 's capabilities---we are not code-limited in considering higher spins and larger mass ratios.  
These runs demonstrate that \dendrogr\ has crossed a key technical threshold, as it is able to access a broader portion of the high-$q$, high-$\chi$ region of parameter space.

\section{Methods for large mass ratios} \label{sec:methods}

Numerical relativity faces a formidable challenge in meeting accuracy requirements for parameter estimation (PE) sufficient for third-generation gravitational wave detectors. 
Moreover, current PE for high mass ratio binaries with
spin for LIGO data analysis pipelines can produce highly inaccurate estimates, as detailed above.
This section highlights the particular numerical methods and algorithms 
that we find essential in performing our current high mass ratio BBH simulations. 
We relegate other details of these simulations to Appendix~\ref{sec:other}.


\subsection{WAMR: Wavelet Adaptive Multi-Resolution} \label{sec:WAMR}

The multi-scale nature of the binary black hole problem makes 
adaptive mesh refinement essential in  numerical relativity.
\dendrogr\ uses wavelet-based adaptive multi-resolution (WAMR) 
to capture sharp features on an unstructured octree grid \citep{Paolucci1, Paolucci2, DeBuhr+18, Fernando+23}.  
Each of the 24 BSSN fields is expanded
in an interpolating wavelet basis with coefficients analogous to 
Fourier coefficients.  However,
in contrast to a Fourier-like basis, the wavelet basis is local, with compact support.
In addition, the wavelet expansion itself generates the computational domain.
We label the grid points on refinement
level $k$ as $\bfx_k$, where refinement increases with index $k$. 
The numerical solution on $\bfx_k$ is $u_k(\bfx_k)$, which contains all solutions 
$u_j(\bfx_j)$ for $j\leq k$.
The wavelet coefficient, $c_k$, for a particular field
takes the difference of the solution on $\bfx_k$ with the solution on $\bfx_{k-1}$ 
as interpolated onto $\bfx_k$. Mathematically, this can be written 
\begin{equation}
c_k(\bfx_k) = u_k(\bfx_k) - \mathscr{P}\left[ u_{k-1}(\bfx_{k-1}) \right],
\end{equation}
where the subtrahend represents the solution $u_{k-1}(\bfx_{k-1})$
as interpolated onto $\bfx_k$ using polynomial interpolation.
Refinement triggers if the wavelet coefficients
exceed a user-defined threshold, $\left| c_k(\bfx_k) \right| > \epsilon_\psi(\bfx_k)$, 
referred to as the wavelet tolerance. 
We then only coarsen once errors fall a fixed factor below the wavelet tolerance; 
this buffer between refinement and coarsening ameliorates 
vacillation in octree level between each refinement pass. 
We find that a smaller factor is optimal prior to merger 
due to the rapid grid changes as the BHs move through spacetime.  
After merger, a larger factor is preferred, as the grid becomes 
less dynamic and is primarily propagating the outgoing gravitational wave (GW).
Finally, we emphasize that in our tests it is 
important to refine on \emph{all} 24 BSSN variables.  
While it might seem reasonable that refining on only a few fields
is sufficient to resolve the black holes and their dynamics, 
we find that all evolved fields---even those without an immediate physical interpretation
---must be adequately resolved 
to obtain high quality results. 

Similar to $p$-adaptivity in finite element codes, 
WAMR ensures no features above a given amplitude go unresolved on the grid; 
features below the target threshold, where $\left| c_k(\bfx_k) \right| < \epsilon_\psi(\bfx_k)$, are considered irrelevant.  
By increasing the order of the representation---i.e., by including additional
terms in the wavelet expansion $c_{k+1}$---we \textit{also} increase the depth of the grid to $\bfx_{k+1}$, doubling the resolution. 
Together, this gives WAMR a built-in convergence akin to both $h$- and $p$-adaptivity. 
This means WAMR's choice of refinement naturally converges toward the true solution as wavelet tolerance decreases (and the maximum refinement depth increases). 
WAMR can thus resolve all locally smooth features simulated on the grid, 
with convergence controlled by a single knob: 
the wavelet tolerance $\epsilon_\psi$. 

We follow the causal WAMR prescription of \citet{Black+25}, in which we
smoothly relax from a maximum wavelet tolerance $\epsilon_{\psi}$ (listed in Table~\ref{tab:runs}) to a minimum tolerance $\epsilon_{\min} = 10^{-5}$, 
but only in regions causally connected 
to the GW emission far from the wake of the gauge wave. 
This reduces over-refining on spurious back-reflections of the gauge wave as it heads toward lower refinement.

\subsection{Onion: Refinement floor about the BHs} \label{sec:onion}

While WAMR itself can in principle handle all refinement on the grid, 
there are situations in which added guidance can lower constraint violations and improve computational efficiency. 
For example, we must constrain WAMR to a maximum level of refinement or it will pursue an arbitrary number of refinement levels at the puncture, which would clearly be unacceptable.  
We thus set a maximum number of permissible refinement levels.  
Another situation where hard-coded requirements are beneficial to guide the refinement involves the organization of the refinement levels in the neighborhoods of the black holes.  Managed well, adding hard-coded refinement  
decreases constraint violations generated from grid changes. 
The nested refinement about each black hole is dynamic 
as the black holes move across the grid.  
As a result, some regions of finest WAMR refinement 
can incur small changes in error that lead to boundary 
points moving in and out of different refinement levels.
Each change in the refinement invokes new interpolations, 
which can generate additional Hamiltonian constraint violation on the grid. 
Refinement therefore must strike a balance between frugality and regularity. 

To this end, we set a sphere of maximum refinement centered about each of the punctures out to radius $R_{\rm AMR}$. 
This helps reduce the stress of repeated refinement and coarsening 
near the apparent horizon and ensures a set refinement level
to a prescribed coordinate extent about each puncture. 
If we let this sphere coarsen as fast as the octree's 2:1 balancing condition permits~\cite{Sundar+08},
then nearby regions of the grid can develop dramatically different resolutions.  This can lead to severe back-reflections, particularly of the initial gauge wave at early times. 
We therefore only permit the sphere to coarsen by $n$ levels
after we reach a radius $(\gamma_{\rm AMR})^n \, R_{\rm AMR}$ away from the punctures, where $\gamma_{\rm AMR} \geq 1$. 
This is a refinement \emph{floor}, not a mandated coarsening. 
Initially, we keep the AMR ratio at $\gamma_{\rm AMR} = 2$.  
Doing so maintains a large region of nested refinement levels---a sort of ``onion" structure---around each BH.
This structure mitigates short lived multi-level changes in refinement as, for example, a sharp feature of the early time dynamics propagates outward. 
Afterwards, we decrease this AMR ratio to the golden ratio 
$\gamma_{\rm AMR} \doteq 1.618$, thereby reducing necessary refinement
(cf. the radial extents of the concentric circles about each puncture in the upper and lower frames of Figure~\ref{fig:frames},
growing far more compact as we transition $\gamma_{\rm AMR} \doteq 2 \to 1.618$ from the upper panel to the lower). 
The current code version parametrizes this as a retarded-in-time sigmoid transition, 
smoothly evolving from the large 2:1 spheres
to a much more localized onion structure about each BH. 
This significantly reduces the overall grid size for large mass ratios, commensurately improving speed and efficiency of these runs.  
These hard-coded onion-like refinement structures about each BH, while not strictly necessary, 
reduce the stress of refinement changes on the evolution system,
leading to lower constraint violations,
especially at large mass ratios.

\subsection{ORBIT: Causal refinement floor} \label{sec:ORBIT}  

High mass ratio inspirals generate higher-order multipole moments 
with amplitudes close to that of the dominant $(\ell, m) = (2, 2)$ mode \citep{Varma+14, Calderon+16, Harry+18}. 
Constructing the full waveform thus depends heavily on resolving these high-frequency modes. 
By construction, WAMR only refines up to a given amplitude, 
so cleanly resolving low-amplitude GW modes (e.g. as they exponentially decay in the ringdown) requires additional, user-set refinement. 

The ORBIT refinement floor of \citet{Black+25} uses the inspiral frequency of the BHs to Nyquist-resolve GWs of a particular target spin-weighted spherical harmonic order $m_{\rm max}$. 
This is approximately equivalent to using a time-retarded grid spacing of the form 
\begin{equation}
  \Delta x \leq 1/[m_{\max} \, f_{\rm ret}^{(2,2)}], 
\end{equation}
where $f_{\rm ret}^{(2,2)}$ is the retarded frequency of the dominant $(2,2)$ mode,
as estimated from the retarded BH orbital angular frequency $\omega$. 
The error of the $(2,2)$ mode, $\varepsilon^{(2,2)}$, as a function of the max Nyquist-resolved mode $m_{\max}$, is 
\begin{equation}
  \frac{\varepsilon^{(2, 2)}}{\varepsilon_{\rm Nyquist}} = 1 - \frac{1}{\sqrt{1 + 1/m_{\max}^{2}}} 
  \sim \frac{1}{2 \, m_{\max}^2},
\end{equation}
given relative to the error at Nyquist resolution $\varepsilon_{\rm Nyquist}$. 
This gives errors of 1.0\% and 0.3\% for $m_{\max} = \{7, 12\}$. 
Thus the ORBIT refinement floor helps resolve waveforms up to 
a given target spherical harmonic order, 
ensuring the grid is sufficiently well-resolved, 
without requiring a similarly high level of resolution 
in other spatial and temporal regions of the simulation.

\subsection{HD: Hamiltonian constraint damping} \label{sec:HD}

As described in \citet{Black+25} we implement a version of Hamiltonian constraint damping (HD) inspired by \cite{Etienne_24}.  The original idea was to 
include a damping term into the evolution of the conformal factor of 
the form 
\begin{equation}
  \partial_t \chi = [\cdots] + c_{H} \chi \frac{(\Delta x)^2}{\Delta t} \mathcal{H},
  \label{conformal_fac_evo_eq}
\end{equation}
where 
  $[\cdots]$ represents the standard RHS terms in this equation, 
  $\chi = {\rm e}^{-4\phi}$ is the conformal factor, 
  $\Delta x$ gives the local grid spacing, and 
  $\Delta t$ gives the time step taken on that grid (this paper uses a global timestep). 
The derivation of the above assumes local grid spacing $\Delta x \ll 1 \, M$, but at grid edges very coarse regions can have $\Delta x > 1\,M$. 
In the present work we see decreased constraint violations through a slight modification.  
In particular, we replace 
\begin{equation}
  (\Delta x)^2 \to \frac{(\Delta x)^2}{1 + 10 (\Delta x/M)^2}
\end{equation}
which keeps this term significantly smaller than $1 \, M^2$. 
This prevents the added term in Eq.~\ref{conformal_fac_evo_eq} from inadvertently dominating the evolution equation.  
While the structure of this form of HD is similar to that in \citet{Etienne_24}, 
we find that HD can damp $\cal H$ by up to an additional $\sim 20\%$. 
Under global time-stepping, keeping the larger BH stable with this aggressive damping scheme 
requires more resolution about the larger BH at higher mass ratios.

The maximum permissible damping amplitude, $c_{H}$, depends on 
the shape of the time integrator's absolute stability region $\mathcal{S}$ in the complex plane. 
Because HD is dissipative, its damping eigenvalue $\lambda_{\rm HD} \propto c_H$ will lie along the negative real axis. 
To be stable, $c_{H}$ must be small enough that $\lambda_{\rm HD} \cdot \Delta t$ lies within $\mathcal{S}$. 
In RK4, the negative-real intercept of $\mathcal{S}$ is $|r_{\rm RK4}| \doteq 2.8$, resulting in maximum $c_H = 0.06$. 
Other time integrators will require different $c_H$ values.
For example, the multi-step RK4 (MSRK) schemes of \citet{Sanches+26} 
have $|r_{\rm MSRK}| \approx 1.1$--$1.4$, 
roughly halving maximum permissible $c_H$.

\subsection{SSL: Slow-start lapse} \label{sec:SSL}

We employ the ``slow-start lapse'' (SSL) introduced by \citet{Etienne_24}. 
SSL includes an additional term in the evolution equation for the lapse: 
\begin{equation}
  \partial_t \alpha = [\cdots] - \chi^{1/2} \bigl(h_{\rm SSL} \, e^{-\tfrac12 (t/\sigma_{\rm SSL})^2}\bigr)\,(\alpha - \chi^{1/2}) \\
\end{equation}
where, again, $[\cdots]$ represents the usual terms in the lapse evolution equation.   
This additional term briefly drives the lapse toward the initial solution, ameliorating the rapid changes of the gauge wave. 
Without SSL, the gauge wave increases in both amplitude and frequency proportional to the mass ratio of the system. 
As the settling of the gauge wave happens on timescales related to the mass of each BH, one might expect the amplitude and duration of SSL to scale similarly. 

The original work, tuned for mass ratios $1 \le q \leq 6$, used a damping amplitude $h_{\rm SSL} = 0.6 / M$ and duration $\sigma_{\rm SSL} = 20 \, M$ (where $M$ is the total ADM mass of the system). 
We find that using the smaller BH's mass $m_{\min}$ rather than the total mass of the system $M$ improves the behavior of SSL and lowers overall constraint violations. 
We therefore use for our $q=24$ simulations
\begin{align}
  h_{\rm SSL} &= 0.3 / m_{\min}, \quad
  \sigma_{\rm SSL} = 40 \, m_{\min}
\end{align}
(which yields identical SSL parameters for $q=1$). 
As the minimum mass of the $q=24$ simulations is $m_{\min} = 1/25$, this results in damping amplitude $h_{\rm SSL} = 7.5 / M$ and duration $\sigma_{\rm SSL} = 1.6 \, M$. 
This tunes the SSL toward the large gauge wave emitted by the smaller BH. 
Since starting the runs, we have found $h_{\rm SSL} = 0.12/m_{\min}$ and $\sigma_{\rm SSL} = 100 \, m_{\min}$ slightly decreases total constraints, better balancing smaller and larger BH. 
For the $q=12$ simulations we use the original SSL parameters, which yield constraint violations of a similar order of magnitude.

SSL permits us to use more sensitive WAMR at very early times.  
As refining on the gauge wave can be quite costly, this change permits a better refined gauge wave, which, in turn, results in fewer back-reflections and a much cleaner waveform.  
While \citet{Fernando+23} saw gauge waves 
needing refinement until $\sim 250 \, M$, 
the gauge wave in the current simulations only requires refinement 
for $\lesssim 10 \, M$, largely due to SSL. 
Resolving the gauge wave with this early refinement 
reduces waveform noise by an order of magnitude. 

\section{Results}\label{sec:results}

\begin{table*}[!ht]\centering
  \caption{
    Summary of runs featured in this paper; see text for variable definitions. 
    (Parentheticals indicate a run in progress.)  
  }
  \begin{ruledtabular}
  \begin{tabular}{lccccccccclccccc}
    Run Label & $q$ & $D_0/M$ & $2m_1/dx_1$ & $2m_2/dx_2$ & $N_1$ & $N_2$ & $\vec \chi_1$ & $\vec \chi_2$ & $m_{\max}$ & \; $\epsilon_{\psi}$ & $N_{\rm orbits}$ & $t_{\rm CPU}$ (hours) & $t_{\rm wall}$ (days) \\
    \hline 
    $\epsilon_{\psi} = 10^{-3}$ & 24 & 8 & 128 & 85 & -- & -- & $0$ & $0$ & 12 & $10^{-3}$ & 11.1 & $5.63 \times 10^6$ & 94.3 \\ 
    $\epsilon_{\psi} = 10^{-4}$ & 24 & 7 & 128 & 85 & 191 & 65 & $0$ & $0$ & 12 & $10^{-4}$ & (4.6) & ($2.18 \times 10^6$) & (41.1) \\ 
    $\chi = 0.0$, lo  & 12 & 8 &  57 & 76 & 82 & 60 & $0$ & $0$ & 7 & $10^{-4}$ & 7.8 & $664 \times 10^3$ & 22.3 \\ 
    $\chi = 0.0$, hi  & 12 & 8 & 113 & 76 & 165 & 60 & $0$ & $0$ & 7 & $10^{-4.5}$ & 7.0 & $807 \times 10^3$ & 22.0 \\ 
    $\chi = 0.4$, AA  & 12 & 8 & 113 & 76 & 141 & 55 & $-0.4\, \hat{z}$ & $+0.4\,\hat{z}$ & 7 & $10^{-4}$ & 4.0 & $510 \times 10^3$ & 13.0 \\ 
    $\chi = 0.8$, AA  & 12 & 8 & 113 & 76 & 81 & 38 & $-0.8 \, \hat{z}$ & $+0.8 \, \hat{z}$ & 7 & $10^{-4}$ & 1.8 & $315 \times 10^3$ & 8.2 \\
    $\chi = 0.8$, AL  & 12 & 8 & 113 & 76 & 81 & 37 & $-0.8 \, \hat{z}$ & $-0.8 \, \hat{z}$ & 7 & $10^{-4}$ & 1.7 & $280 \times 10^3$ & 7.3 \\
    $\chi = 0.8$, PR  & 12 & 8 & 113 & 76 & 81 & 38 & $+0.8 \, \hat{x}$ & $-0.8 \, \hat{z}$ & 7 & $10^{-4}$ & 4.1 & $737 \times 10^3$ & 18.8 \\
  \end{tabular}
  \end{ruledtabular}
  \label{tab:runs}
\end{table*}

\begin{figure*}[t]\centering
  \includegraphics[width=.49\linewidth]{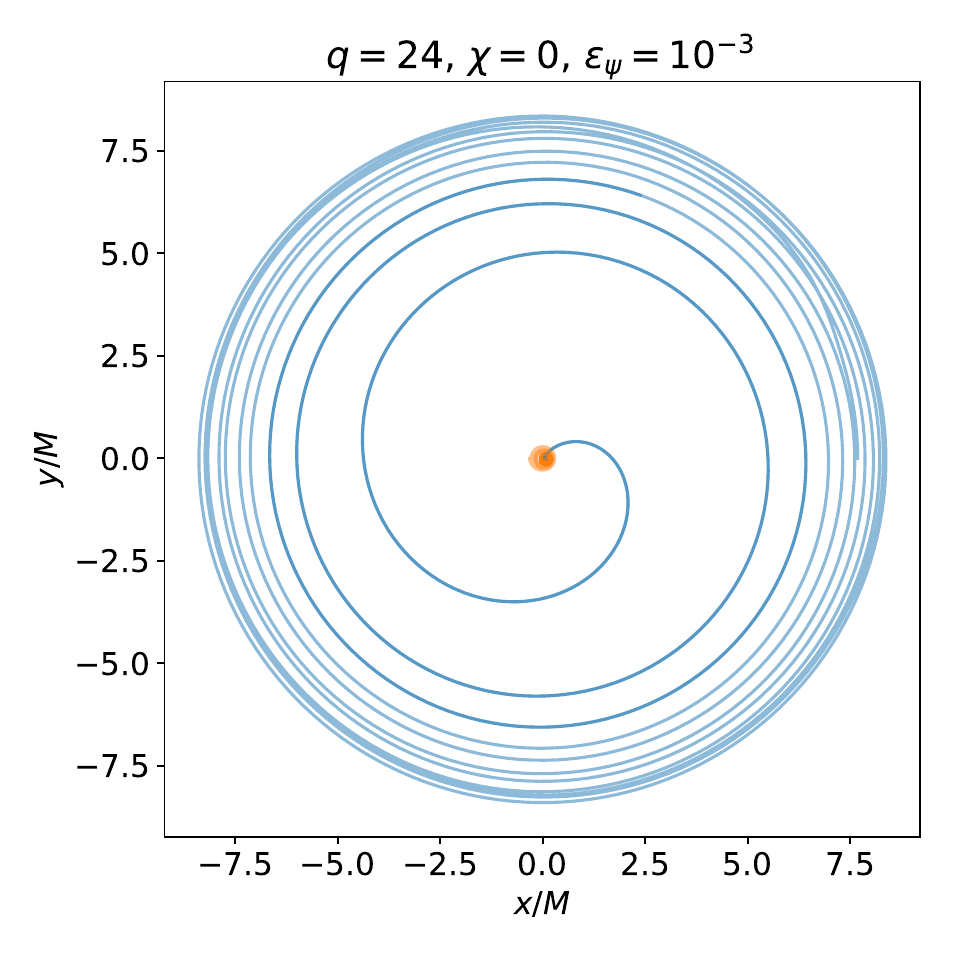}
  \includegraphics[width=.49\linewidth]{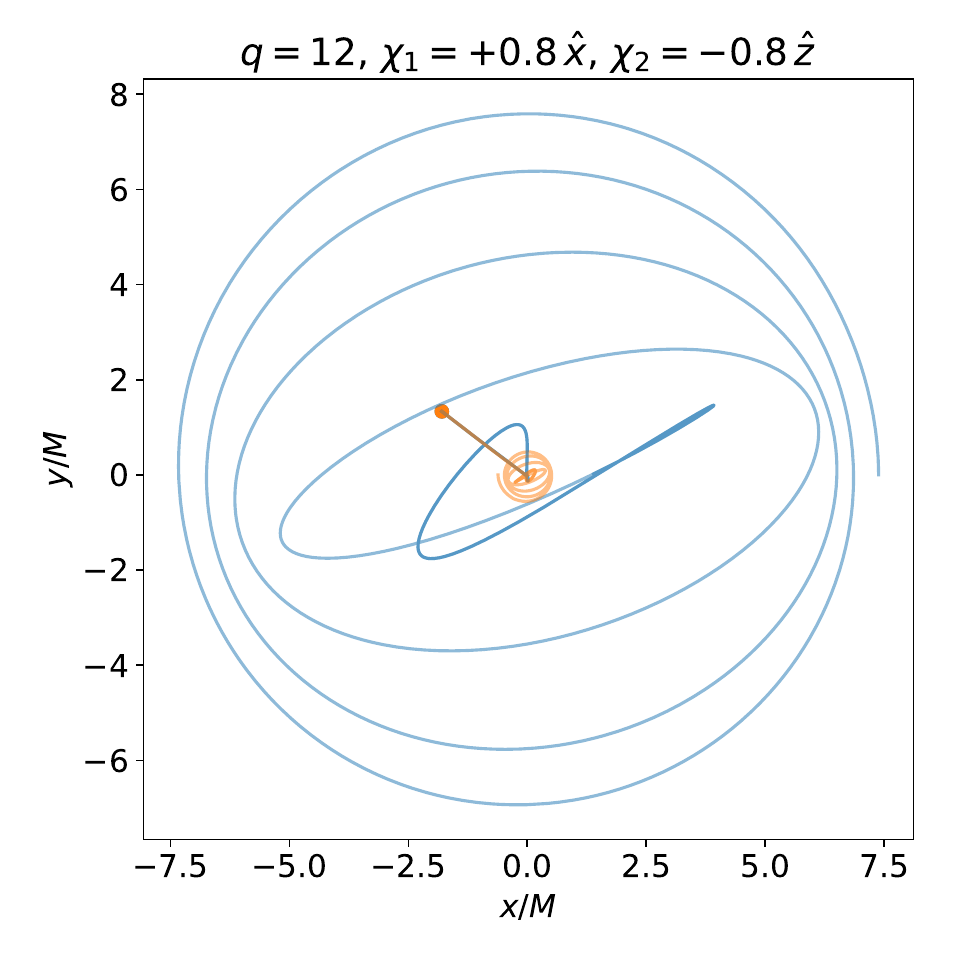}
\caption{
    Orbital trajectories for the frontier \dendrogr\ binary black hole inspirals. 
    Coordinate paths (integrated from the shift) shown for 
      $q=24$ (nonspinning;
      wavelet tolerance $\epsilon_{\psi} = 10^{-3}$) on the left, while 
      the right panel shows $q=12$ with large spins on each BH: 
        $\chi_1 = +0.8 \hat x$ for the larger BH and
        $\chi_2 = -0.8 \hat z$ for the smaller BH (highly precessing). 
    Paths colored by mass (orange as primary, blue as secondary); line width scaled by $z$-distance. 
  }
  \label{fig:paths}
\end{figure*}

We present the first \dendrogr\ simulations in the high-mass-ratio 
$\times$ high-spin regime that LISA and other 3G detectors will probe. 
These simulations drove development of robust numerical techniques 
as outlined in Section~\ref{sec:methods}, 
enabling stable, long-duration evolutions at high mass ratios and large spins. 
In particular, we consider $q=24$ (nonspinning) binary black hole inspirals 
and $q=12$ binaries with spins up to $\chi=0.8$, 
including a highly precessing configuration. 
We expect the utility of our methods to extend to yet higher-$q$ binaries with large spin; 
these runs reflect not the limits of our current capabilities, 
but merely the extent of our most recent tests. 

Our major purpose in this paper is to discuss numerical techniques for large mass-ratios,
not necessarily to deliver catalog-ready, fully converged waveforms. 
As such, we do not perform multi-resolution convergence studies here. 
As we developed our methods, several runs had short inspirals, 
with significant gauge wave noise infiltrating the waveforms. 
Subsequent work will produce production quality waveforms, 
suitable for astrophysical parameter estimation. 
Significantly, no simulations crash at merger, 
  all demonstrate low and stable constraint violations, 
  and each of the well-resolved simulations achieves excellent mass conservation. 
  See Appendix~\ref{sec:stability} for more details. 
We are unaware of any code issues 
which would prevent us from pursuing even higher mass ratios with similarly complex spin configurations. 
This paper demonstrates that \dendrogr\ 
can push further into this regime of parameter space and 
begin to systematically produce high-$q$ $\times$ high-$\chi$ simulations.

\subsection{Simulation parameters} \label{sec:parameters}
\vspace{-.5cm}

Table~\ref{tab:runs} presents parameters 
and summary data for each simulation. 
We focus here on mass ratios $q = 12$ and $24$. 
We begin all simulations at a nearby separation distance of $D_0 = 8\,M$ 
with the exception of the high-resolution $q=24$ run, 
which started at $D_0 = 7\,M$. 

The table lists an approximate point count across the horizon as $2 m_i / dx_i$,
where $dx_i$ is the resolution at the punctures relative to the ADM mass $m_i$ of each black hole. 
Here $i=1$ is the primary (larger BH) while $i=2$ is the secondary. 
Using the BH\-aH\-AHA-measured mean apparent horizon radius, $r_i$, we calculate 
$N_i \equiv {\rm med}(2 r_i / dx_i)$, 
the simulation's median point count across each horizon. 
Note that in our gauge, coordinate horizon radii do not scale precisely with BH mass;
smaller BHs generally have $r_2/m_2 < 1$ whereas
larger BHs generally have $r_1/m_1 > 1$, so
this ratio will differ between the two BHs. 
Furthermore, the current HD formulation requires 
increased resolution about the larger BH for stability. 

These test runs have dimensionless spin vectors $\vec \chi_{i}$
in the $\hat{z}$-direction with the exception of the precessing run. 
Here ``AA'' indicates spin--spin anti-alignment, ``AL'' indicates aligned spins, and ``PR'' indicates precessing orientations. 
All AA \& AL runs are retrograde, with the smaller BH orbiting against the angular momentum of the larger BH. 
ORBIT ensures refinement Nyquist-resolves up to at least 
order $m_{\max} = 12$ for the two $q=24$ runs 
and $m_{\max} = 7$ for each of the $q=12$ runs. 
We use the causal refinement structure of \citet{Black+25} 
with minimum wavelet tolerance $10^{-5}$ and 
maximum wavelet tolerance $\epsilon_{\psi}$
spanning $10^{-3}$ (low resolution) to $10^{-4.5}$ (high resolution). 

Finally, we present the number of orbits to merger $N_{\rm orbits}$ \footnote{
  We find that the time of maximum relative velocity 
  coincides closely with common horizon formation time;
  we therefore use it as a proxy for merge time,
  as not all cases had a fully functional implementation of BH\-aH\-AHA. 
}, 
the total CPU hours spent on completing the run, $t_{\rm CPU}$, 
and the total wall time spent on completing the run, $t_{\rm wall}$. 
Both times are measured $200 \, M$ past merger 
for completed runs.

\vspace{-.5cm}
\subsection{Pioneering runs}
\vspace{-.25cm}

\begin{figure*}\centering
  \includegraphics[width=\linewidth]{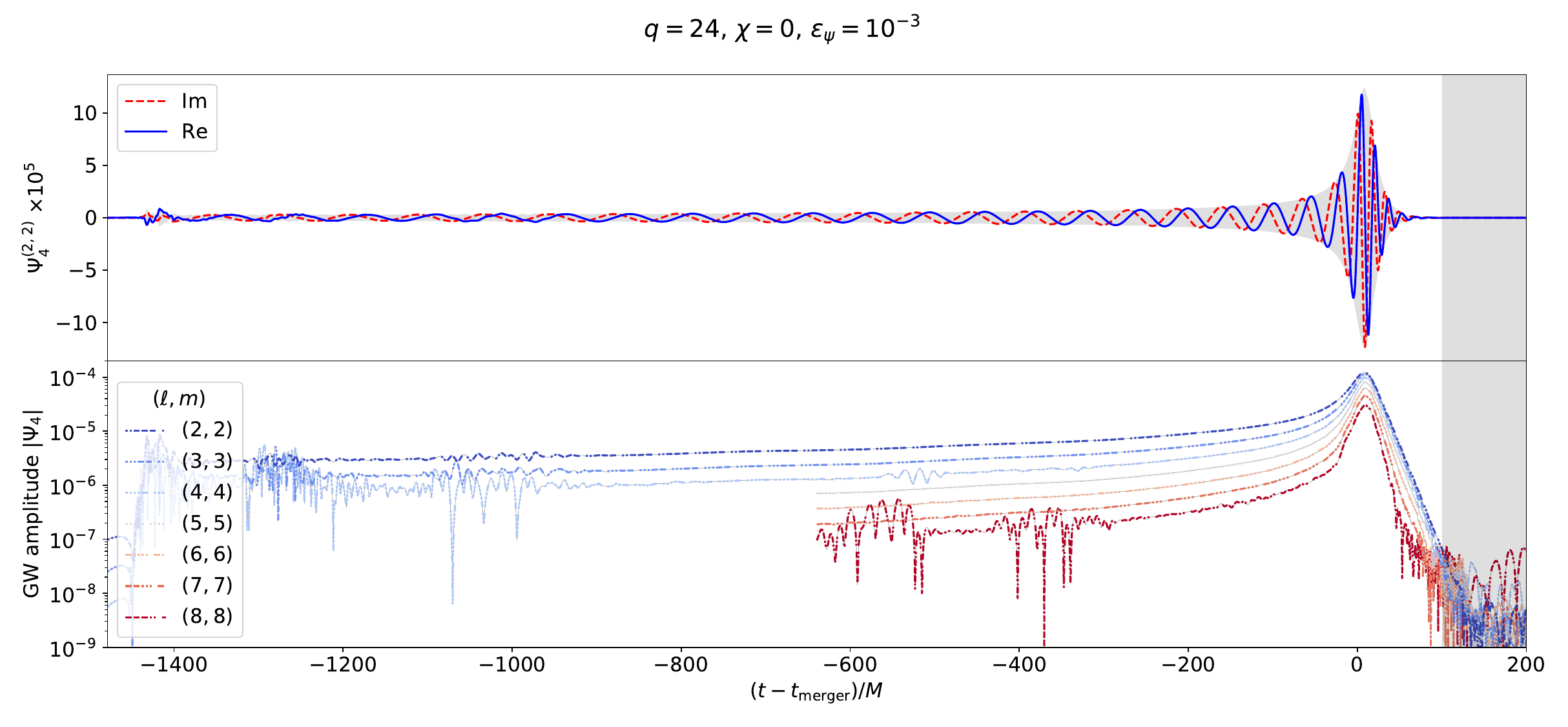}
  \caption{
    Gravitational waveforms for the mass ratio $q=24$ run
    with wavelet tolerance $\epsilon_{\psi} = 10^{-3}$. 
    \textbf{Upper panel:} Spin-weighted spherical harmonic mode $\Psi_4^{(2,2)}$ 
    colored by real part (blue solid), imaginary part (red dashed), 
    and total magnitude (grey background). 
    \textbf{Lower panel:} 
    Waveform amplitudes for spin-weighted spherical harmonic modes $\ell = m$ up to $\ell = 8$. 
    Higher-degree modes $\ell > 4$ were only enabled later in the evolutions (output only; refinement was unaffected). 
  }
  \label{fig:q24_wave_and_amp}
\end{figure*}

\begin{figure*}\centering
  \includegraphics[width=\linewidth]{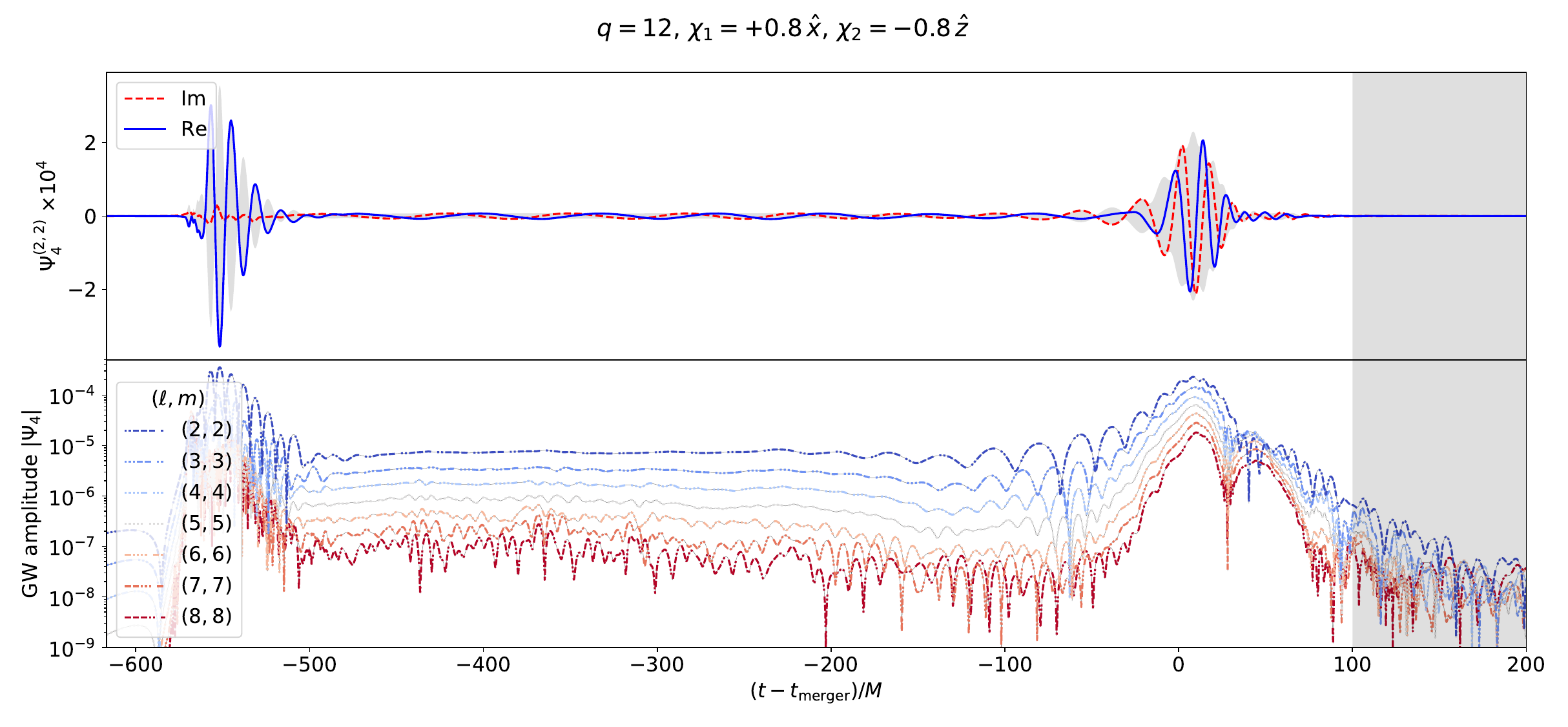}
  \caption{
    As Figure~\ref{fig:q24_wave_and_amp}, but for the $q = 12$, $\chi = 0.8$, highly precessing run. 
  }
  \label{fig:q12PR_wave_and_amp}
\end{figure*}

This suite of runs pushes the boundaries of our previously explored parameter space. 
In this paper, we move from our previously largest mass ratio of $q=16$  to $q=24$ (with no code limitations to our knowledge preventing us from going higher). 
We also, for the first time, publish runs with significant spin magnitude, up to $\chi = 0.8$, including a highly precessing case. 
In the following, we highlight several features of our simulation with largest mass ratio
and our simulation with the most complex spin configuration.

Each of the simulations successfully ran through merger 
and beyond or are on their way toward doing so. 
No code errors led to crashes in these simulations.  
To the best of our knowledge, we are not algorithm-limited 
in progressing to more extreme configurations. 
We discuss validation measures in Appendix~\ref{sec:stability},
including constraint violations, horizon mass conservation, and other accuracy indicators.

\subsubsection{Coordinate trajectories}
\vspace{-.25cm}

The coordinate trajectories for the $q=24$ run and the precessing $q=12$
runs are shown in Figure~\ref{fig:paths}; 
BH positions are shift-integrated. 
The left panel shows the circular inspiral of 
our $q=24$ nonspinning binary (wavelet tolerance $\epsilon_{\psi} = 10^{-3}$); 
the right panel depicts our $q=12$ highly-precessing case. 
The runs successfully and stably reach merger and beyond. 
See the appendices for more information on coordinate trajectories, 
including retrograde path reversal near merger (Appendix~\ref{sec:path_reversal}) 
and estimated kick velocities post-merger (Appendix~\ref{sec:kick_velocities}).

\subsubsection{Waveforms}
\vspace{-.25cm}

We extract spin-weighted spherical harmonics of the Weyl tensor $\Psi_4^{(\ell, m)}$ 
at radius $R_{\rm GW} = 50 \, M$ to highlight 
waveform noise otherwise less visible at larger radii. 
Figures~\ref{fig:q24_wave_and_amp} \&~\ref{fig:q12PR_wave_and_amp} 
show real and imaginary portions of the $(\ell, m) = (2, 2)$ mode 
alongside an amplitude plot of all $\ell = m$ extracted modes for the
$q=24$ and precessing $q=12$ binaries, respectively.
For a complete depiction of all waveforms generated, 
see Appendix~\ref{apx:waveform_gallery}.

The amplitudes of these runs follow the expected mode hierarchy; 
even up to multipole  
$\ell = 8$ 
we see strong structure and clarity in mean behavior 
(especially after the initial ringing from the gauge wave and its echoes fade). 
The secular behavior through inspiral, merger and ringdown 
is consistent with expectations. 
As will be shown in \S\ref{sec:discuss_WAMR}, 
decreasing the wavelet tolerance (i.e., forcing additional WAMR-based refinement) 
decreases waveform noise dramatically.

The precessing $q=12$ waveform shows the characteristic 
amplitude and phase modulation driven by spin--orbit coupling, 
particularly once the system is within $200 \, M$ of merger
(at which point the majority of the velocity points toward $\hat{z}$; 
the orbital inclination moves to $>45^\circ$ above the $xy$-plane). 
As the binary leaves the $xy$-plane, 
power transfers from $\ell = m$ modes toward $\ell \ne m$ modes, 
as expected from the rotation of the orbital plane.

\subsubsection{Comparison to RIT catalog} \label{sec:cf_RIT}
\vspace{-.25cm}

\begin{figure}\centering
  \includegraphics[width=\linewidth]{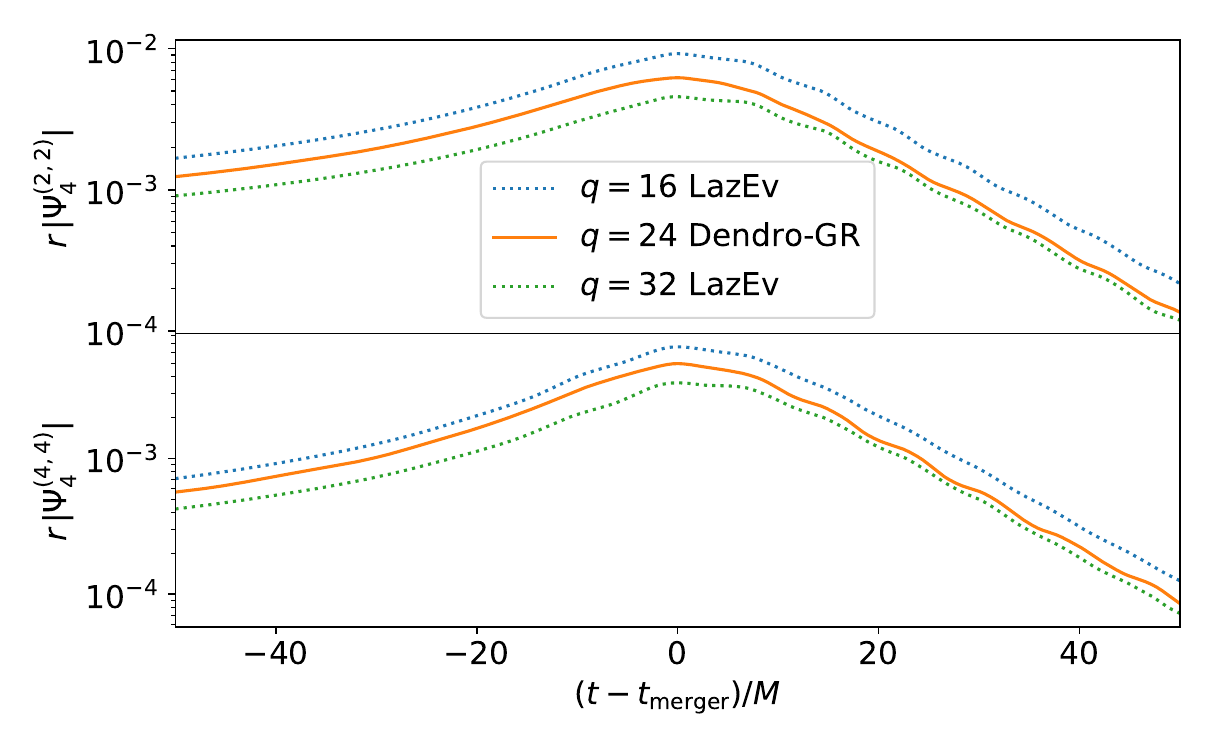}
  \caption{
    Comparison of mode amplitudes between the $q=24$ \dendrogr\ run (at extraction radius $R_{\rm GW} = 50 \, M$) and 
    neighboring $q=16$ \& $q=32$ runs from LazEv simulations (extrapolating $R_{\rm GW} \to \infty$). 
    Upper panel: $(\ell,m) = (2,2)$ mode. 
    Lower panel: $(\ell,m) = (4,4)$ mode. 
    Our $q=24$ run shows comparable behavior, 
    lying between the neighboring LazEv runs. 
  }
  \label{fig:q24_cf_lazev_modes}
\end{figure}

Public gravitational wave catalogs do not currently
contain $q=24$ waveforms for comparison. 
However, the RIT catalog~\citep{Healy_Lousto_2022} 
does have $\ell \leq 4$ waveforms at nearby mass ratios
of $q=16$ and $q=32$ (RIT:BBH:1965 and RIT:BBH:1025 respectively). 
Peak waveform amplitude runs roughly linearly with respect to 
symmetric mass ratio $\nu \equiv q / (1 + q)^2$~\cite{Healy+17},
so we expect these RIT waveform amplitudes 
(extrapolated to infinite extraction radius) 
to bookend our $q=24$ run 
(extracted at radius $R_{\rm GW} = 50 \, M$).
The RIT runs thus set rough bounds on what should be monotonic behavior. 

Figure~\ref{fig:q24_cf_lazev_modes} shows that
our $q=24$ simulation threads the needle between its
neighboring mass ratio runs from the RIT catalog. 
While this first proof-of-concept test of the 
$\epsilon_{\psi} = 10^{-3}$  (low-resolution) run 
shows some modulation in the ringdown amplitude
(as do the RIT waveforms), stricter WAMR in 
the $\epsilon_{\psi} = 10^{-4}$  (high-resolution) run 
reduces noise by over an order of magnitude 
(as will be discussed in \S\ref{sec:discuss_WAMR}). 
We anticipate decreasing wavelet tolerance (along with other code improvements) will substantially clean our future runs.

%

\section{Discussion: Feature-based refinement and accuracy in NR} \label{sec:discuss_WAMR}

It is relatively straightforward to exchange accuracy for run speed and vice versa.  
Clearly, adding refinement to the grid costs more in compute time but improves code accuracy.  Conversely, coarsening the grid decreases compute cost but decreases overall accuracy. 
Truly efficient improvements to codes must somehow improve one without unduly sacrificing the other. 

Over the course of simulating these runs,
we have implemented a number of code improvements
which have significantly improved the capability of \dendrogr. 
In the remainder of this section, we describe a major improvement to our simulations' overall efficiency: the effects of using 
highly localized, feature-based refinement via WAMR as opposed to well-informed guesses or box-in-box or sphere-in-sphere refinement geometries.

\begin{figure*}\centering
  \includegraphics[width=\linewidth]{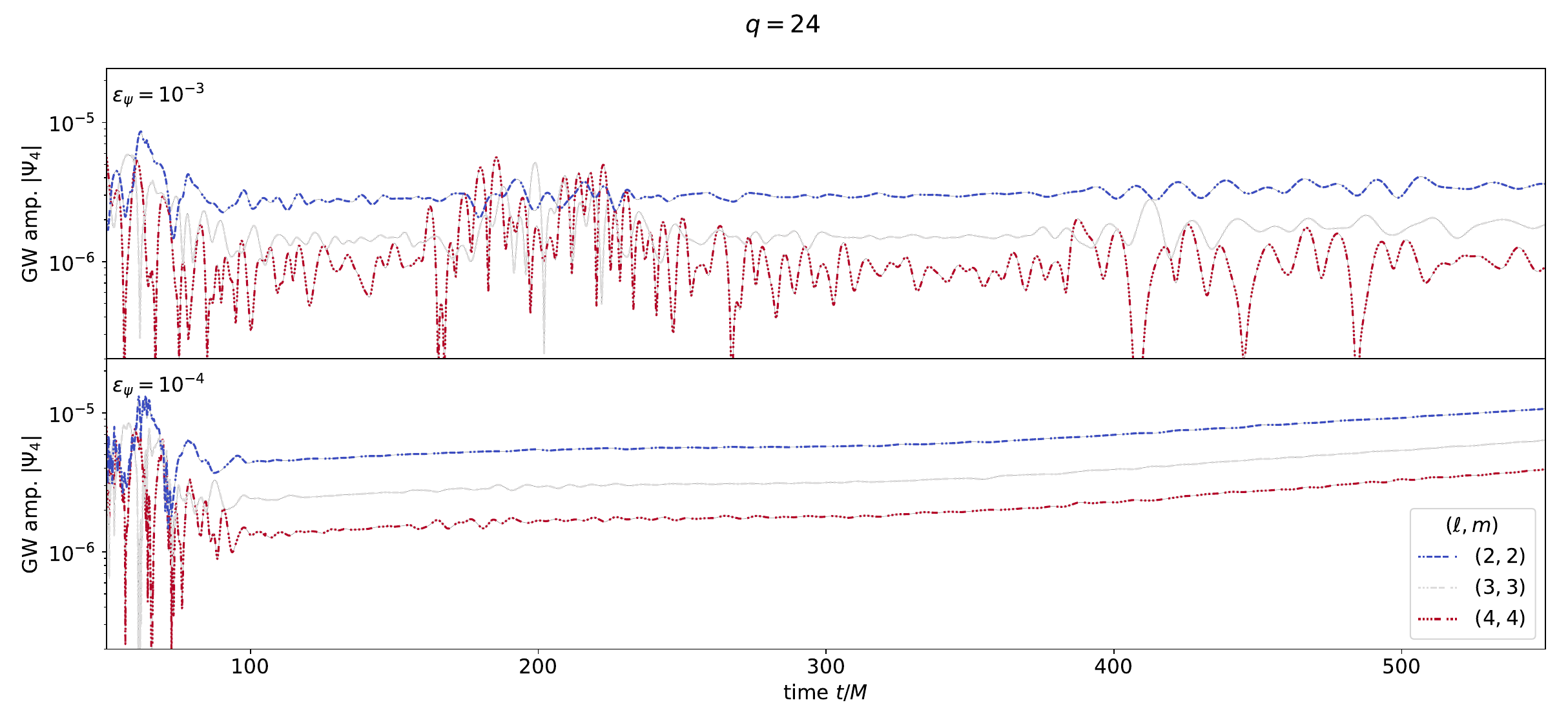}
  \caption{
    Comparison of waveform amplitudes between the two $q=24$ runs. 
    On lowering the wavelet tolerance by an order of magnitude, 
    we decrease waveform noise by an order of magnitude. 
  }
  \label{fig:q24_early_cf}
\end{figure*}

\begin{figure}\centering
  \includegraphics[width=\linewidth]{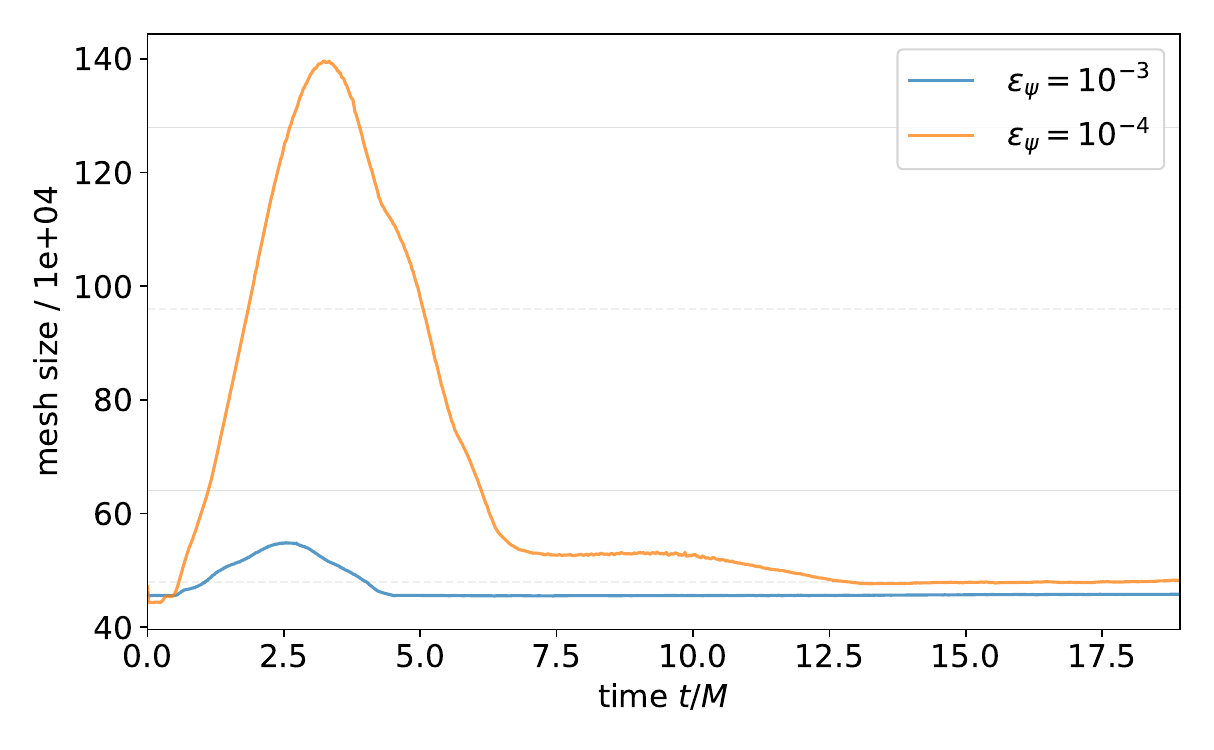}
  \caption{
    The amount of refinement as measured by the size of the mesh (the number of leaf nodes in the octree) near the start of the two $q=24$ runs. 
    The lower wavelet tolerance of the $\epsilon_{\psi} = 10^{-4}$ run
    triggers additional early refinement on the gauge wave, 
    which soon returns to a flat constant near that of the $\epsilon_{\psi} = 10^{-3}$ run. 
  }
  \label{fig:q24_mesh_early}
\end{figure}

\begin{figure}\centering
  \includegraphics[width=\linewidth]{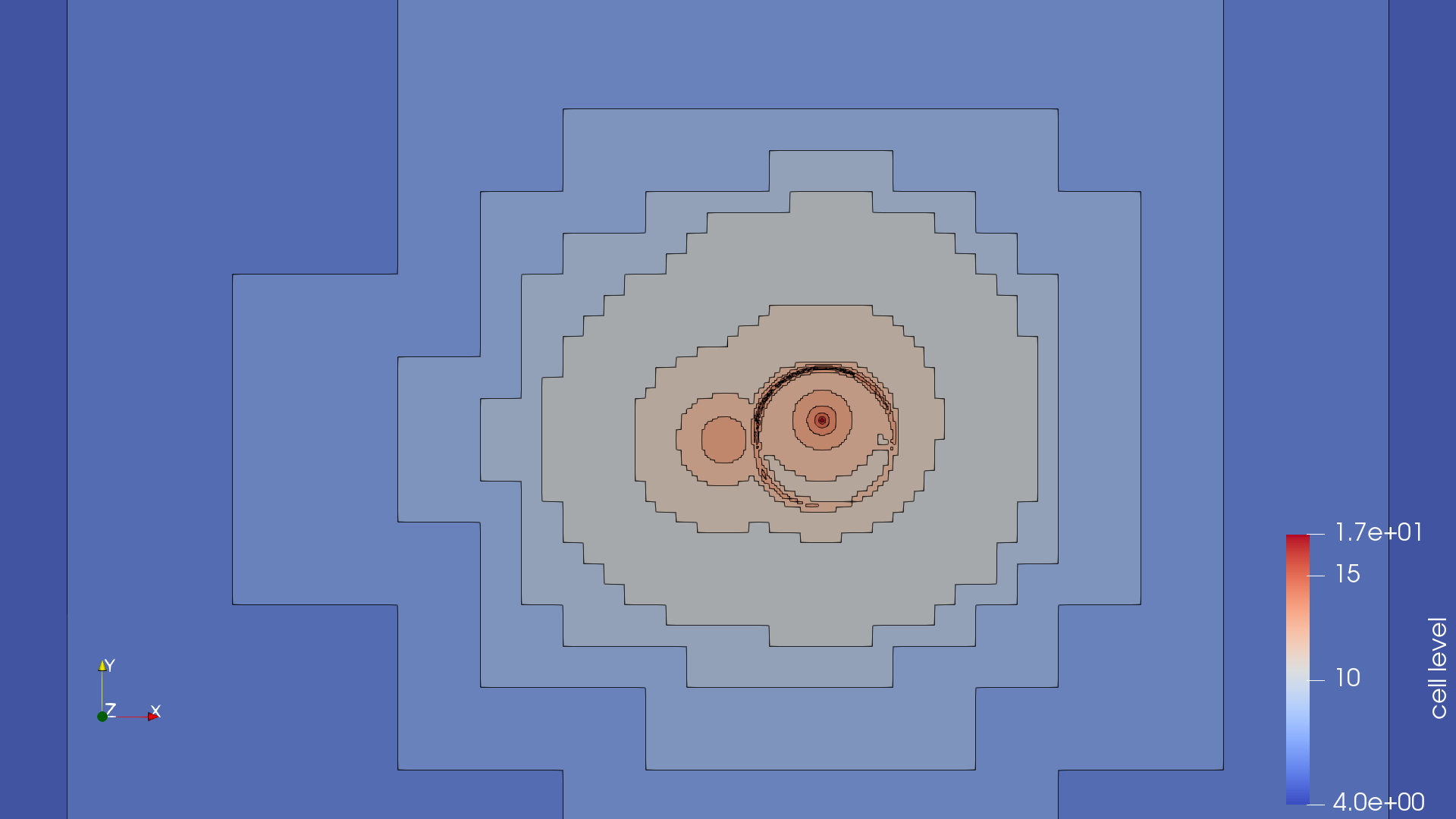}
  \includegraphics[width=\linewidth]{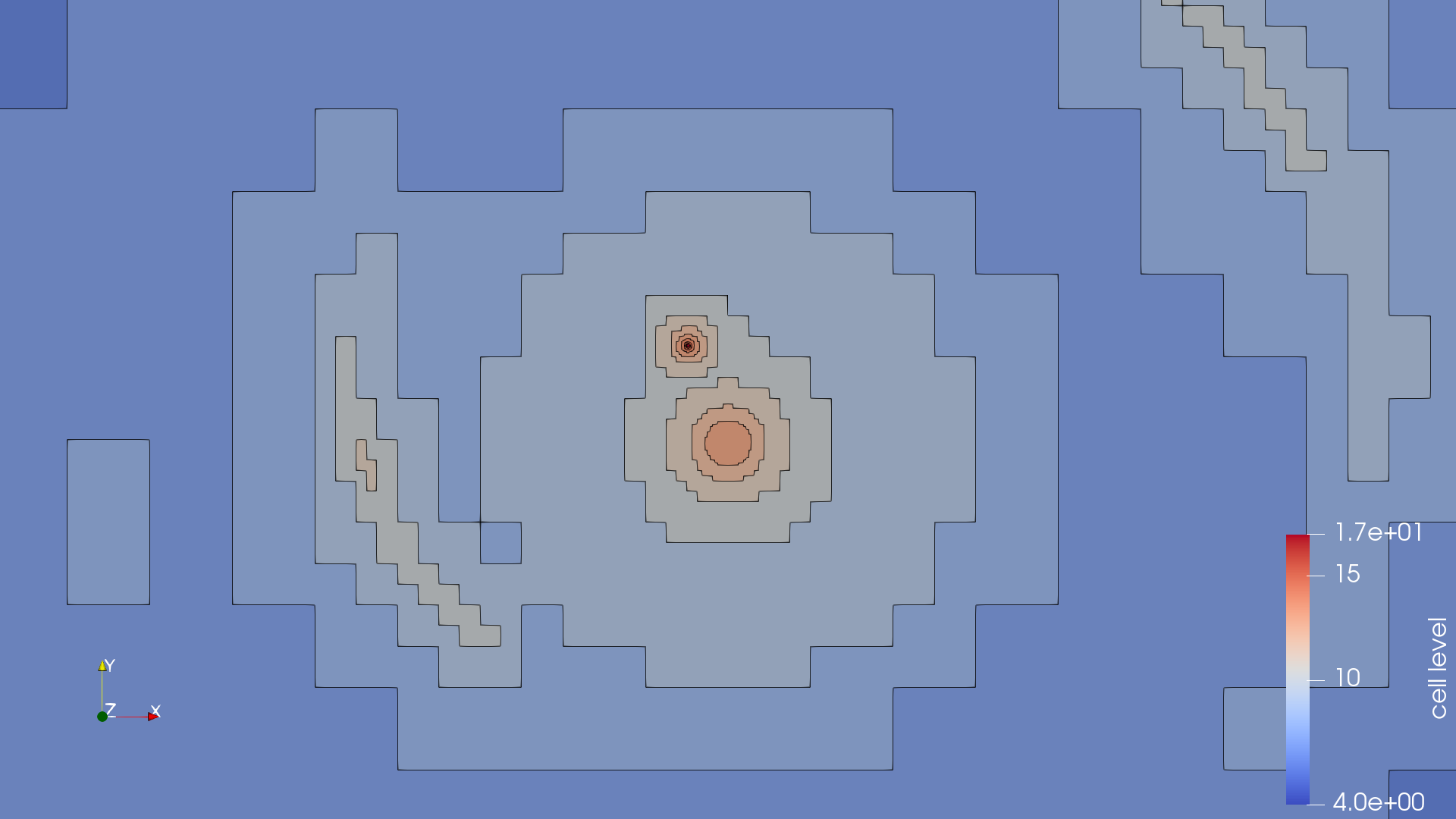}
  \caption{
    Zoom-in on the early time refinement in the high-resolution $q=24$ run ($\epsilon_{\psi} = 10^{-4}$).  The color scheme indicates the depth of refinement in the octree. 
    \textbf{Upper ($t/M \simeq 5$):} WAMR refines heavily on the outgoing gauge wave, aided by the wide onion structure (see \S\ref{sec:onion}), as it escapes the local potential. 
    \textbf{Lower ($t/M \simeq 45$):} Once the gauge wave leaves the local potential, the grid settles down to a more relaxed state, aligned with the low-resolution $q=24$ run ($\epsilon_{\psi} = 10^{-3}$). 
  }
  \label{fig:frames}
\end{figure}

A well known difficulty in moving puncture evolutions is that the
main component of the error at early times is not sourced by the physical binary at all, but rather by the evolving coordinate system.  
In particular, as the conformally-flat initial data relax
toward the two trumpet-like geometries, each puncture emits an
outgoing gauge wave that sweeps across the grid. If this unphysical radiation is insufficiently resolved as it propagates across refinement boundaries, it seeds 
back-reflections and high-frequency noise that persist and manifest in the extracted waveform. 

In \dendrogr, resolving this transient plays a significant role in generating clean waveforms.  
Retaining high resolution around these gauge waves until they have entirely propagated off the grid can be quite costly.  
However, when combined with slow-start lapse (\S\ref{sec:SSL}) and the onion structure of refinement levels around the black holes, the unphysical waves require refinement 
only until they leave the strong-field region of the grid.  
This reduces the time required to follow the gauge
wave from $\sim 250\,M$ (as reported in \citep{Fernando+23}) 
to $\lesssim 10\,M$ \textit{and} reduces the 
waveform noise by an order of magnitude.  
This effect is largely ascribable to these two improvements.  SSL extends the timescale on which the gauge wave is very well resolved.  In conjunction with this, the onion structure prevents rapid, multi-level refinement or coarsening around the black holes, reducing Hamiltonian constraint violation generated by regridding. 
This permits lower wavelet tolerance and therefore extra levels of refinement to follow the gauge waves until they have propagated well beyond the local potential well of the binary, 
at which point the amplitude and need for extra resolution decreases.

To further illustrate the importance of adequately
resolving the gauge wave and the utility of localized
refinement, consider the inspiral waveform 
for two similar runs of a $q=24$ binary.
Figure~\ref{fig:q24_early_cf} compares 
the waveform magnitudes between 
runs with maximum wavelet tolerance 
$\epsilon_{\psi} = 10^{-3}$ (low-resolution) and  
$\epsilon_{\psi} = 10^{-4}$ (high-resolution). 
The total mesh size (the number of leaf nodes
of the octree) for the two runs is
compared in Figure~\ref{fig:q24_mesh_early} 
as a function of grid time.  
The mesh size of both runs sees linear growth initially, 
as they refine to track the gauge wave. 
While the higher-resolution run briefly requires 
a mesh that is three times larger than 
the lower-resolution run, this need
persists for only $t \lesssim 10 \, M$, about the
time required for the gauge wave to scatter away from
the binary's potential well.  
After this time, both runs have comparable mesh sizes.
The higher initial cost of the higher-resolution
run pays dividends in
an order of magnitude reduction in waveform noise
hundreds of $M$ later in the simulation! 
Relative waveform noise decreases by an order of magnitude,
in lockstep with the reduction in wavelet tolerance.

Now consider Figure~\ref{fig:frames} which shows precisely where this  additional refinement
is needed at early times in a binary black hole simulation.
The figure shows the number of cell levels WAMR uses 
to refine on the outgoing gauge wave in the high-resolution $q=24$ simulation.  
Note the thin, roughly spherical shell in the top
frame ($t \simeq 5\, M$) that tracks the gauge
wave expanding away from the smaller mass black hole.
The refinement must follow this expanding and continuously 
moving wavefront until it largely diminishes as it leaves the strong field region.  The bottom frame of 
Figure~\ref{fig:frames} shows the grid after the mesh
has settled at $t \simeq 45\, M$.
This refinement behavior, crucially, is made possible because WAMR excels in localized refinement on 
solution features.
Refinement criteria based on the binary system's
characteristics or guided primarily by intuition, such as fixed nested boxes or
spheres centered on the punctures---a standard moving-box strategy---could not easily follow such an outgoing shell.
Indeed, capturing it geometrically would require refining 
the entire growing spherical volume around each BH, 
and only for the brief window in which the feature exists. 

If one were to develop an \textit{ad hoc} refinement condition for binary
black holes, intuition might guide one to a condition that produces
grids similar to those shown in the bottom frame of Figure~\ref{fig:frames}.  
Such a condition,
however, would miss the crucial refinement needed to propagate the
gauge wave outside of the strong-field region, resulting in noisier waveforms.
Relatedly, the location of the gauge wave 
is not predeterminable.  
The amplitude, frequency, and lifetime of the gauge wave 
scale with the mass ratio and spin~\citep{Etienne_24}, 
and its evolution and scattering depend on the gravitational 
potential.
Hand-tuning similar refinement regions by trial and error 
becomes progressively less tractable in exactly 
the high-$q$, high-$\chi$ regime we target here. 

Wavelet-adaptive refinement addresses this directly and efficiently. 
Because WAMR tags cells based on the local wavelet (interpolation error) 
content of the evolved fields rather than on distance from the punctures
(\S\ref{sec:WAMR}), it detects the gauge wave automatically, refines
precisely the shell it occupies, and coarsens once it has passed---no
prior knowledge of its trajectory is required. 
The same mechanism that supplies the built-in $hp$-adaptivity 
discussed in Appendix~\ref{sec:stability} places resolution where the solution, rather than the coordinate geometry, demands it.

\section{Conclusions} \label{sec:conclusions}

As gravitational wave detectors continue to improve, new gravitational wave sources will
be discovered, adding to our understanding of black hole evolution in our universe.
Waveform models used for data analysis will also need to both improve and be computed
for a wider range of sources to maximize the scientific impact of the new detectors.  
Recent work has shown that existing
data analysis models, developed for nearly equal mass binaries, do not extend well to 
larger mass ratios, $q \gtrsim 10$, for spinning binaries.  More high-quality 
numerical relativity solutions are needed for these high-$q$ and high-spin configurations.

This paper presents \dendrogr\ simulations of binaries with large mass ratios
($q=24$ and $q=12$) and high spins ($q=12$; $\chi \leq 0.8$), including a strongly precessing configuration.  
Our completed runs are stable through ringdown (the $q=24$, $\epsilon_\psi=10^{-4}$ run was still in progress but stable to date), and 
we see a well-behaved mode hierarchy in $\Psi_4$ of the extracted modes (up to $\ell=8$).
Amplitudes across inspiral--merger--ringdown of the $q=24$ run align with the RIT catalog, 
lying neatly between neighboring $q=16$ and $q=32$ runs.

As mentioned, calculating solutions to the full Einstein equations for binary
inspirals with large $q$ and spin remains a challenging computational task.  
A combination of numerical techniques will be necessary to explore this 
region.  This work highlights some promising numerical methods
using \dendrogr.
First and foremost is efficient mesh adaptivity and refinement.
This includes our built-in refinement scheme WAMR, as well as 
the ORBIT approach to causal refinement and the onion-like structure for refinement in the vicinity of the BHs.  
These techniques help to optimize refinement in relation to 
binary black holes and the large mass ratio problem in particular.  
WAMR generates an unstructured grid using
a measure of local convergence of the solution and a wavelet tolerance $\epsilon_\psi$.
Lowering the wavelet tolerance improves the solution
analogous to $hp$-adaptivity in finite elements.
This is demonstrated, for example, in the $q=24$ runs, 
where increasing WAMR sensitivity reduced the waveform noise by an order of magnitude.
Second, SSL and Hamiltonian damping are essential for reducing 
noise in the computational domain.

Pursuing the challenge of high mass ratio, high spin systems calls for significant innovation. 
Extreme gauge waves, tight resolution requirements, 
long inspirals, and significant scale differences
have pushed the design and development of \dendrogr, and they continue to 
drive new innovation in our code. 
First tests with some multi-step Runge--Kutta schemes~\citep{Sanches+26}, which reuse solutions from previous time steps to reduce RHS evaluations, show promising speedups when paired with a reduced $c_H$. 
Likewise, compact finite differences (CFDs) show great promise in increasing code efficiency. 
At the same stencil size, we can achieve order-of-magnitude improvements in accuracy~\cite{Kim_24}. 
Initial tests of CFDs in \dendrogr\ show that these gains carry over
to black hole evolutions with moving punctures~\cite{Garey_26_APS}.
Changes in supercomputing hardware to heterogeneous architectures offer
performance gains at the cost of additional programming complexity.
We are moving \dendrogr\ to a new Python/JAX infrastructure in order to enable the
performance gains of these new architectures, while developing in higher-level
programming languages.
Finally, the use of innovative coordinates adapted to black hole binary geometries
has proven to be extremely useful for both significantly reducing the 
computational problem size and producing highly accurate waveforms~\cite{Ruchlin_18,Etienne_24}.
Our new Python/JAX-based \dendrogr\ can use multiple coordinate patches with
independent octree hierarchies.
The capabilities of \dendrogr\ presented here, combined with the 
new features under active development, make \dendrogr\ well-positioned
to attack large mass-ratio binaries in numerical relativity.

\section*{Acknowledgments} 

This research was supported by NSF grants \href{https://www.nsf.gov/awardsearch/show-award/?AWD_ID=2207615}{PHY-2207615} (BYU) and 
\href{https://www.nsf.gov/awardsearch/show-award/?AWD_ID=2207616}{PHY-2207616} (Utah).
WKB was supported by the College of 
{Computational, Mathematical, and Physical Sciences} 
at Brigham Young University. 
All runs were completed on \href{https://tacc.utexas.edu/systems/stampede3/}{Stampede3} through compute time awarded by
\href{https://access-ci.org}{ACCESS-CI}. 
We thank our collaborators Yosef Zlochower and Chloe Malinowski for many discussions throughout this work. 
%

%

\section*{Data Availability} \label{sec:data_avail}

The version of \dendrogr\ used to produce the results 
in this paper is distributed with the MIT license at 
\href{https://github.com/paralab/Dendro-GR}{github.com/paralab/Dendro-GR}. 

All input parameter files and simulation output for this paper's runs are archived in a Zenodo repository at \href{https://doi.org/10.5281/zenodo.24169963}{doi:10.5281/zenodo.24169963} \citep{Zenodo_Black_2026}. 
Plots were visualized using \href{https://bitbucket.org/wkblack/dendro-gr-analysis-tools/src/master/}{wkblack/dendro-gr-analysis-tools} on Bitbucket.

%

\appendix

\section{Other methods} \label{sec:other}

We use the open-source code \href{https://github.com/paralab/Dendro-GR/}{\dendrogr} for all the simulations in this paper. 
Our current code builds on the structures introduced in \citet{Fernando+23} and \citet{Black+25},
with several modifications described in \S\ref{sec:methods}. 
We evolve the BSSN equations using the standard moving puncture gauge with several modifications. 
We incorporate a version of Hamiltonian constraint damping (see \S\ref{sec:HD}) directly into the evolution of the conformal factor which significantly decreases constraint violations. 
We also add a tuned ``slow-start lapse" \citep{Etienne_24}
to the evolution equation for the lapse (see \S\ref{sec:SSL}). 
Following the prescription of \citet{Fernando+23}, 
we focus the shift-damping parameter $\eta$ centrally, 
using $\eta \approx 2/M$ within the orbital region, smoothly transitioning at $r = 40 \, M$ 
to $\eta \approx 0.25/M$ in the wave zone. 
We experimented with other shift-damping terms, especially those that
localize and tune the damping about the individual black holes, such as
the $\eta_{\rm G}$ prescription introduced by \citet{Healy_Lousto_2022}.
However, these shift-damping terms triggered spurious refinement with WAMR; 
we obtained better results using a constant value for $\eta$ in the strong-field region.

Additional parameters used for these runs: 
\begin{itemize}
  \item Refinement-sphere radius $R_{\rm AMR} \doteq 1.55\,m_i$ for the $q=24$ runs and $R_{\rm AMR} \doteq 2\,m_i$ for the $q=12$ runs. 
  \item $\gamma_{\rm AMR}$ transition ($2 \to 1.618$): for most runs this was applied via a manual restart; only the $q=12$, $\chi=0$ hi run used a sigmoid in $\tau = t - r/\sqrt{2}$, with width $5 \, M$ and midpoint $t_0 = 50 \, M$. The precise timing and shape of this transition had minimal material effect on waveforms. 
  \item Pre- and post-merger coarsening factors: \\ $0.2$ and $0.03125$, respectively.
  \item Courant factor ${\rm CFL} = 0.25$. 
\end{itemize}

We generate initial data using the {\sc TwoPunctures} initial data solver \citep{Ansorg+04}. We determine initial momenta via the \textsc{NRPyPN} quasi-circular orbit solver \citep{NRPyPN}, yielding low-eccentricity configurations. 
While all our first derivatives are 6th order,
the $q=24$ runs inadvertently used 4th-order second-derivative schemes.
This increased constraint violations by a factor of 
two to five above what our code typically achieves, 
but did not materially affect the results presented here.

\section{Waveform gallery} \label{apx:waveform_gallery}

\begin{figure*}\centering
  \includegraphics[width=\linewidth]{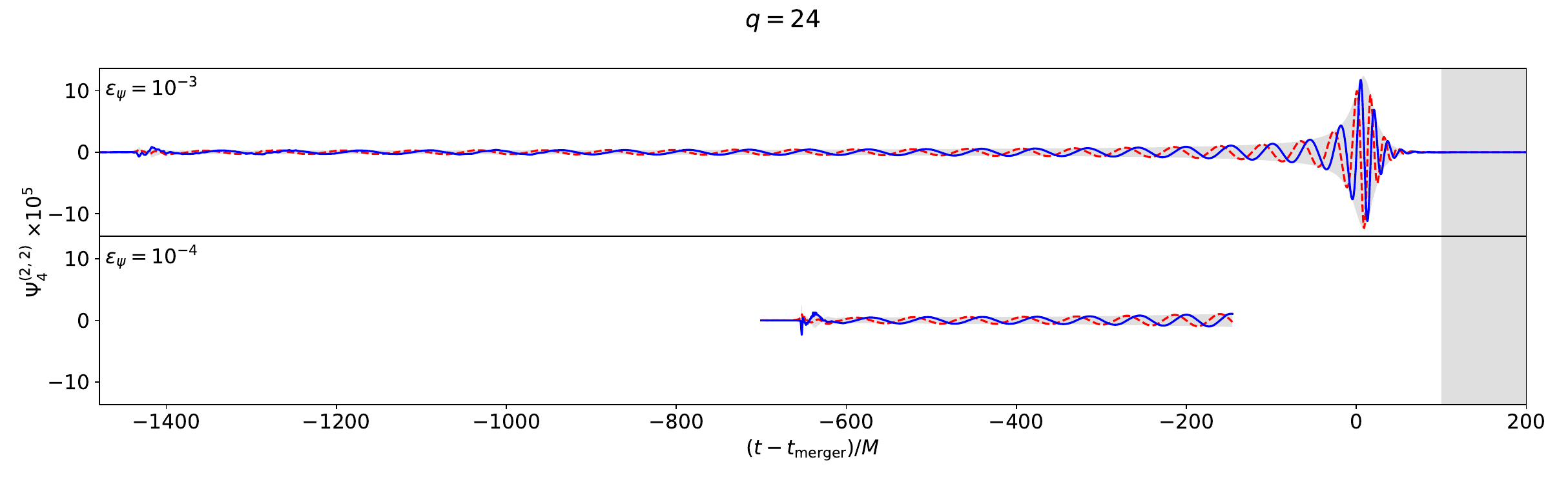}
  \caption{
    Gravitational waveforms for the two $q=24$ runs, 
    started at initial separation $D_0 = 8 \, M$ (upper) 
    and $D_0 = 7 \, M$ (lower), aligned at merger. 
    (For the unfinished high-res run, we match final coordinate separation to the low-res run.)
    Spin-weighted spherical harmonic mode $\Psi_4^{(2,2)}$ 
    colored by real part (blue solid), imaginary part (red dashed), 
    and total magnitude (grey background). 
  }
  \label{fig:q24_wave}
\end{figure*}

\begin{figure*}\centering
  \includegraphics[width=\linewidth]{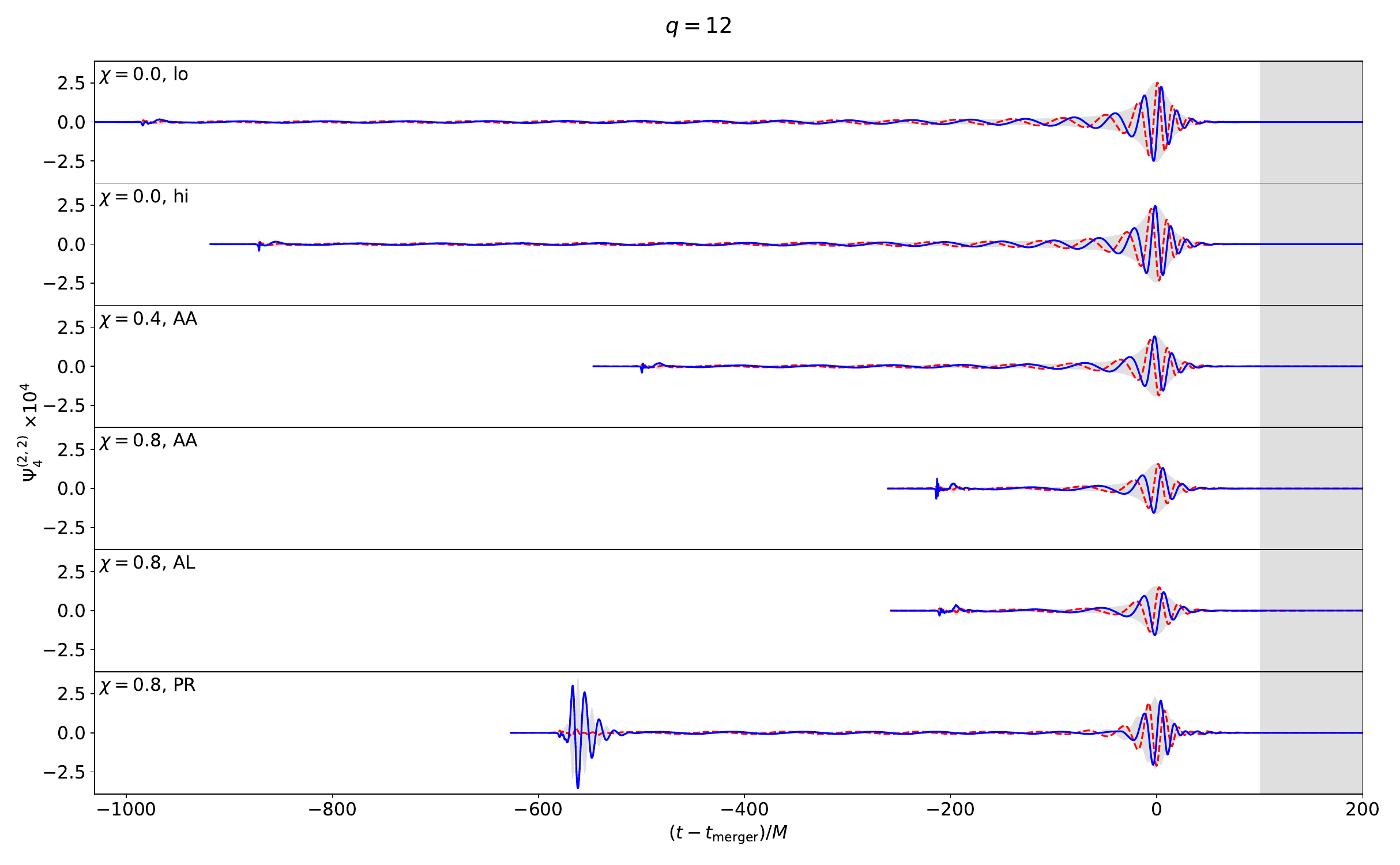}
  \caption{
    As Figure~\ref{fig:q24_wave} but for the $q=12$ runs. 
  }
  \label{fig:q12_wave}
\end{figure*}

We here display waveforms for all runs, aligned at merger. 
We extract spin-weighted spherical harmonic 
decompositions $(\ell, m)$ of the Weyl scalar $\Psi_4$ 
at several radii, spanning here from $50 \, M$ to $100 \, M$. 
Displaying plots at the relatively small radius of $50 \, M$
makes gauge effects and waveform noise more visible, 
providing more stringent tests of our code. 
Figure~\ref{fig:q24_wave} shows the waveforms for the two
$q=24$ runs with the estimated merger time set to $t=0$.
The suite of $q=12$ waveforms are shown in Figure~\ref{fig:q12_wave}.

\begin{figure*}\centering
  \includegraphics[width=\linewidth]{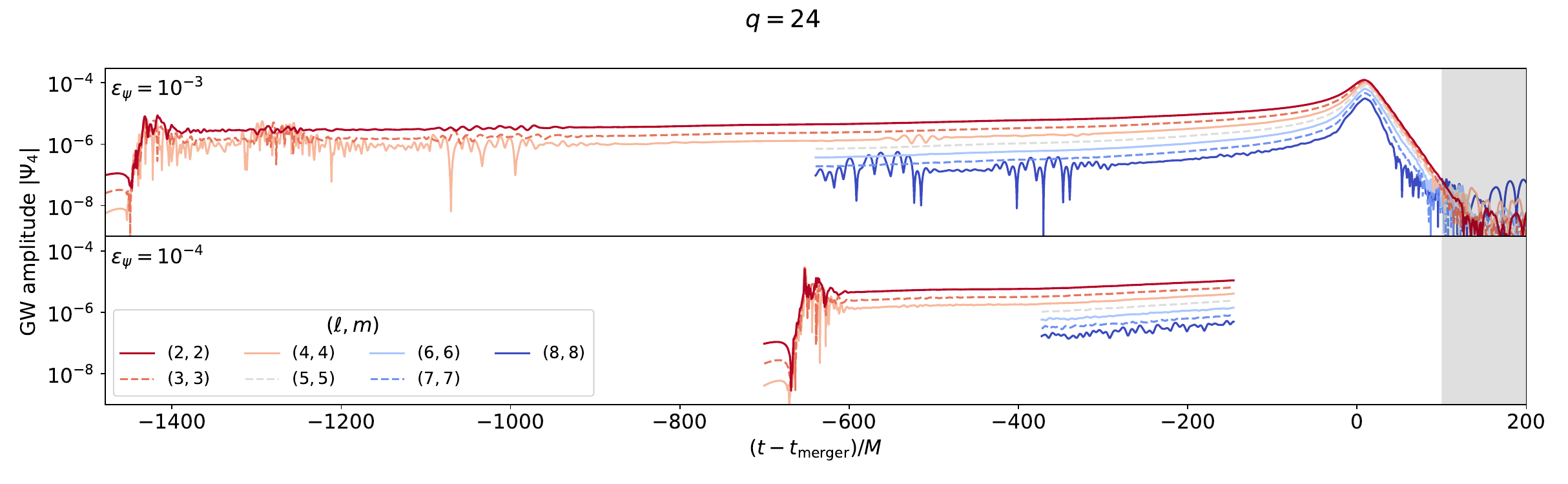}
  \caption{
    Waveform amplitudes for both $q=24$ runs of Figure~\ref{fig:q24_wave}. 
    Only spin-weighted spherical harmonic modes $\ell = m$ up to $\ell = 8$ are shown here for clarity of plotting. 
    Higher-order modes $\ell > 4$ were only enabled later in the evolutions (output only; refinement was unaffected). 
  }
  \label{fig:q24_wave_amp}
\end{figure*}

\begin{figure*}\centering
  \includegraphics[width=\linewidth]{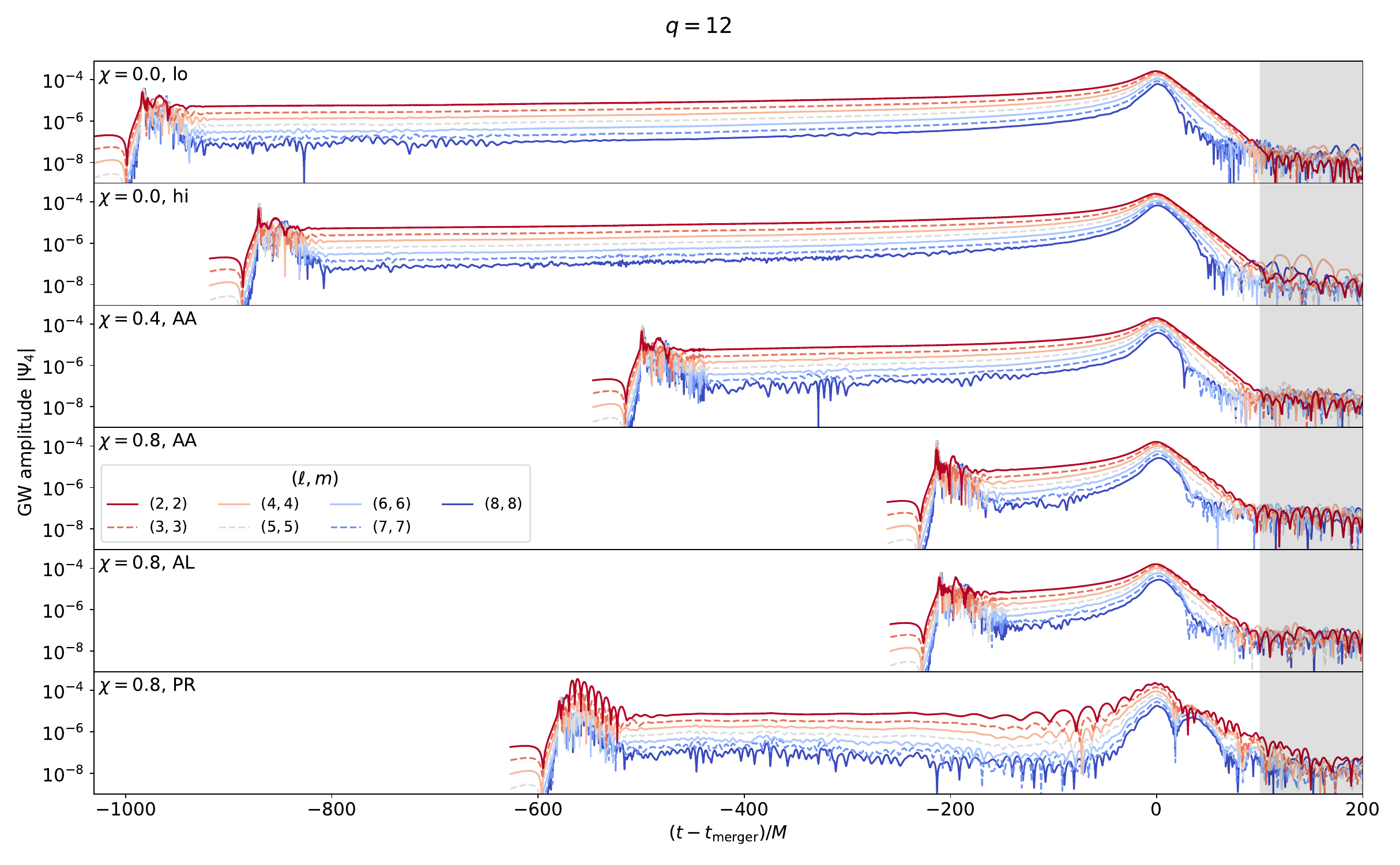}
  \caption{
    As Figure~\ref{fig:q24_wave_amp} but for the $q=12$ runs. 
  }
  \label{fig:q12_wave_amp}
\end{figure*}

The secular portions of the waveforms in Figure~\ref{fig:q24_wave_amp}
show the expected hierarchy of structure. 
For the $q=24$, $\epsilon_{\psi} = 10^{-3}$ run, noise remains elevated during
the first $\sim 500\,M$, visible even in the $(2,2)$ mode. 
The otherwise equivalent $\epsilon_{\psi} = 10^{-4}$ run with 
increased WAMR sensitivity has far cleaner modes, 
with even the $(5, 5)$ mode appearing clean at a glance. 
(See \S\ref{sec:discuss_WAMR} for more on why this change was effective.) 
While the amplitude of the $(8,8)$ mode is often noisy on short timescales, 
its average, smoothed behavior follows the expected mode hierarchy for waveforms.

The ringdowns for the $q=12$ and $\chi \leq 0.4$ runs are largely well-behaved, 
as shown in Figure~\ref{fig:q12_wave_amp},
following the expected exponential ringdown decay at low $\ell$,
where grid support and signal strength are greatest. 
However, the $\chi = 0.8$ runs show significant beating of amplitude,
comparable to that seen in similar LazEv waveforms. 
We leave the origin of these features to future investigation.

\section{Simulation quality metrics} \label{sec:stability}

This section presents several measures to establish 
the validity of our BBH merger solutions.  
These measures include the convergence of the wavelet expansion,
stability of constraint violations, 
conservation of horizon mass, and waveform quality.
The following sub-sections provide direct 
evidence that \dendrogr\ simulates these 
systems well, even when the secondary is 
24 times smaller than the primary, or when
high spins drive strong frame-dragging.

\subsection{Wavelet multi-resolution convergence} \label{sec:wamr_convergence_appendix}

First, we note that WAMR automatically measures the local
convergence of the solution as it generates the multi-resolution grid.
By construction, the wavelet coefficients remain below 
the set tolerance $\epsilon_\psi$ everywhere on the grid. 
The combination of expanding the wavelet representation and
generating finer grid points is analogous to 
$hp$-convergence. (See \S\ref{sec:WAMR} for more discussion.)
As shown in \S\ref{sec:discuss_WAMR}, as we dial down the wavelet tolerance, 
the simulation indicates convergence toward its true solution.

\subsection{Constraint violation stability} \label{sec:cvs}

As a diagnostic, \dendrogr\ evaluates the Hamiltonian and momentum constraints
at all grid points outside small exclusion regions around each BH.  These regions are scaled by the individual BH masses and have radii $R_{\rm excl} \sim 1.55 \, m_i$.  This allows the punctures and their immediate neighborhoods to be excluded while still including much of the strong-field region.  
We then combine them in quadrature into 
the total constraint violation $C_{\rm TOT}$. 

This measure for evaluating the constraints (averaging across leaf nodes of the octree) differs from typical volume-weighted calculations of constraint violations. 
Because leaf-weighted constraint violations give greater weight to regions of greater refinement, they provide more information about regions of the grid with more significant dynamics and suppress somewhat more smooth outer regions of the grid. 
Leaf-weighted constraint violations are less sensitive to changes in simulation box size, focusing attention in the strong-field region. 
This makes the magnitude of $C_{\rm TOT}$ considerably larger 
(by $\sim 10^{3.5} \times$) than volume-weighted constraint violations. 

\begin{figure}\centering
  \includegraphics[width=\linewidth]{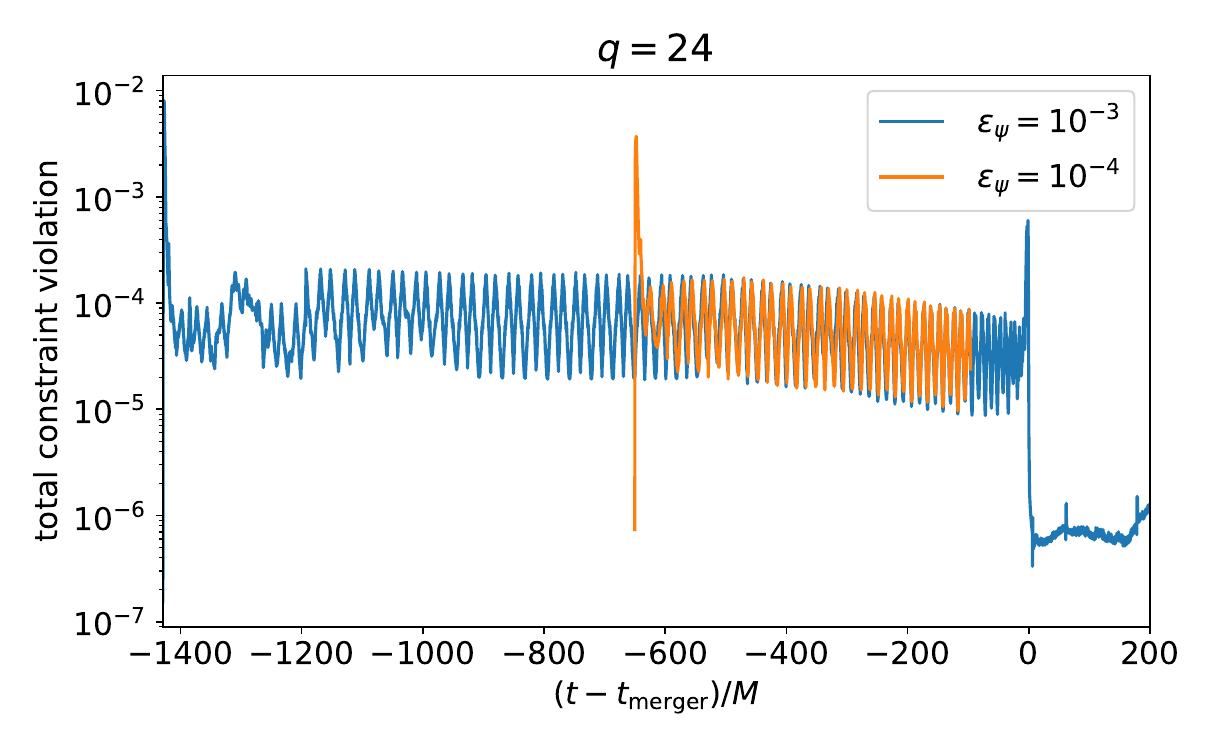}
  \caption{
    Stable total constraint violation $C_{\rm TOT}$ at leaf nodes for both $q=24$ runs,
    remaining low throughout the simulation.
    $C_{\rm TOT}$ is the leaf-averaged RMS of $\mathcal{H}$ and $\mathcal{M}_i$ (in quadrature) over all grid points outside a small exclusion region about each BH
    (see \S\ref{sec:cvs}).
  }
  \label{fig:q24_cvs}
\end{figure}

\begin{figure}\centering
  \includegraphics[width=\linewidth]{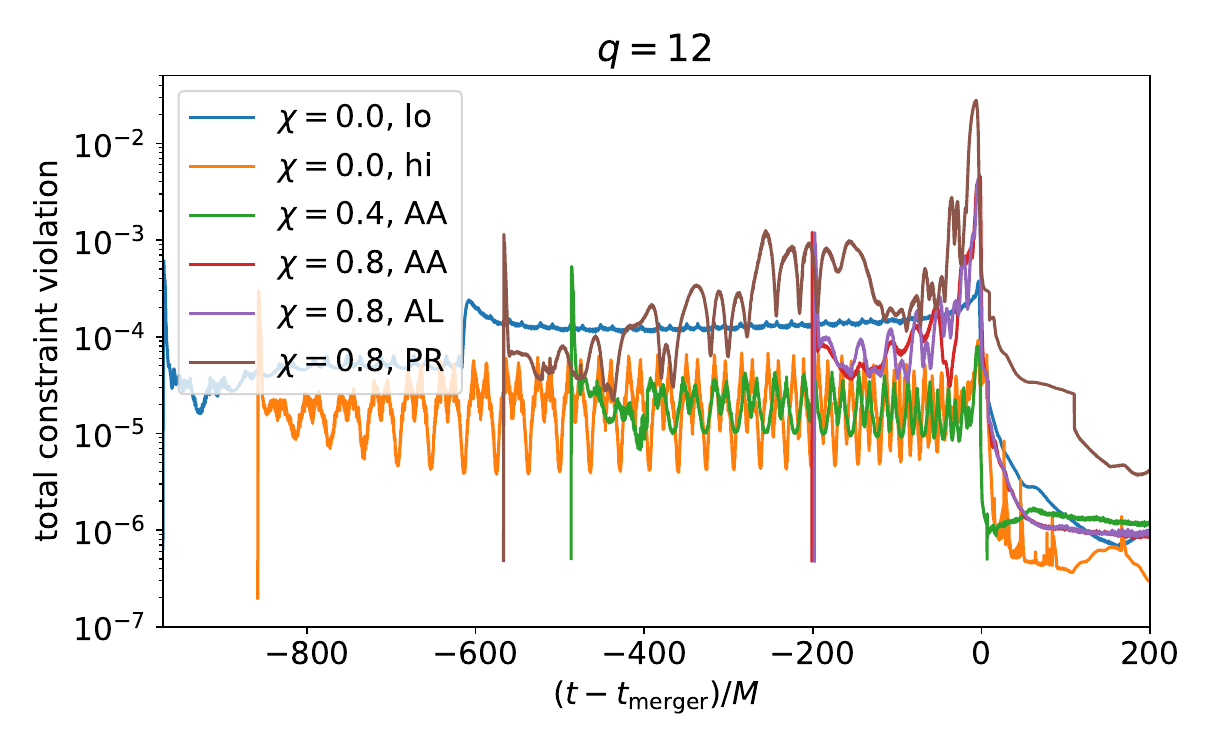}
  \caption{
    As Figure~\ref{fig:q24_cvs} but for the $q=12$ runs.
  }
  \label{fig:q12_cvs}
\end{figure}

The Hamiltonian and momentum constraint violations
remain bounded and largely constant across each simulation, with no runaway growth. 
Figure~\ref{fig:q24_cvs} shows that for $q=24$ average constraint violations at leaf nodes remain low ($\lesssim 10^{-4}$) throughout the simulation. 
Besides brief spikes initially and at merger, the constraints are stable or decreasing throughout both runs. 
Figure~\ref{fig:q12_cvs} shows that for the $q=12$ runs constraint violations typically stay between $10^{-5}$ and $10^{-3}$. 
The high-spin runs show a larger increase during inspiral and significant but brief spikes at merger, followed by rapid decay. 
The high-resolution re-run of the spinless case 
shows constraint violations approximately $10 \times$ lower overall, 
with the Hamiltonian constraint periodically reduced by factors of $\sim 50$. 

\subsection{Horizon mass conservation} \label{sec:BHaHAHA}

\begin{figure}\centering
  \includegraphics[width=\linewidth]{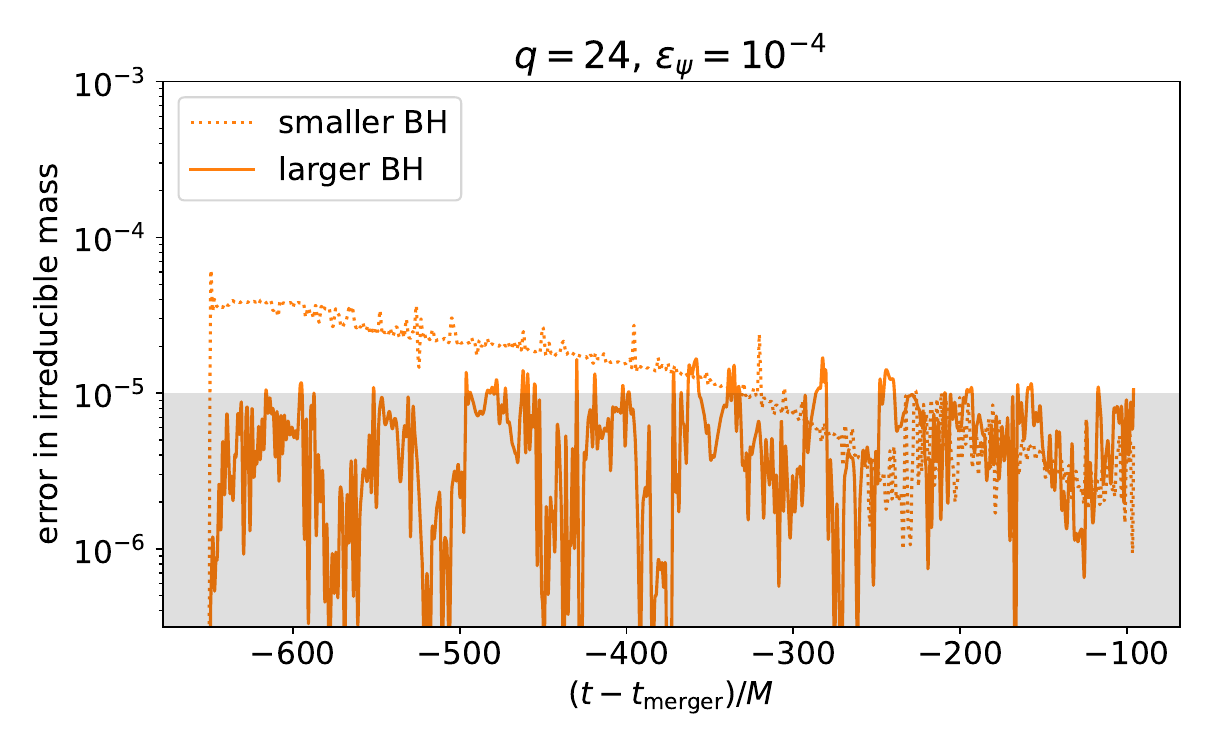}
  \caption{
    Relative accuracy of apparent horizon masses for
    the $q=24$ run at wavelet tolerance $\epsilon_{\psi} = 10^{-4}$
    as measured by BH\-aH\-AHA.
    Error relative to target irreducible mass as given
    by Eqn.~\eqref{eqn:m_irr} from initial data input.
    Dotted lines indicate the smaller BH
    while solid lines represent the larger BH.
  }
  \label{fig:q24_bah}
\end{figure}

\begin{figure}\centering
  \includegraphics[width=\linewidth]{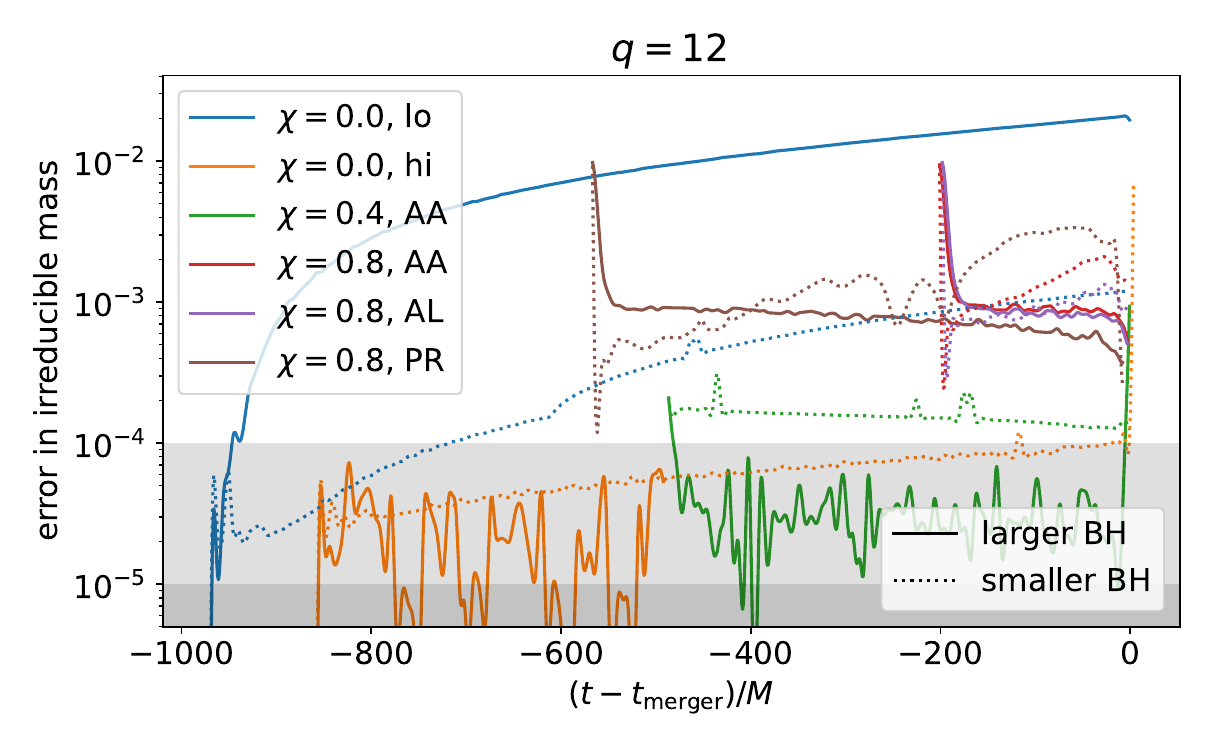}
  \caption{
    As Figure~\ref{fig:q24_bah} but for each of the $q=12$ black holes.
  }
  \label{fig:q12_bah}
\end{figure}

After the longest run ($q=24$, $\epsilon_{\psi} = 10^{-3}$) began, we added the BH\-aH\-AHA\ 
apparent horizon finder~\citep{Etienne+25} to \dendrogr. 
BH\-aH\-AHA provides fast and accurate measures of the apparent horizons for our most recent simulations. 
BH\-aH\-AHA\
measures horizon circumferences in each Cartesian plane to estimate the BH spin. 
The BHs take some time to settle into their secular horizon shapes, 
so the first $t \sim 20 \, m_i$ 
(where $m_i$ is the mass of the individual BH) 
of the spin estimates exhibits significant ringing \footnote{
  This ringing seems to follow quasinormal mode (QNM) ringing timescales: 
  $\tau_{\rm decay} \geq 11.241 \, m_i$, 
  diverging as spin approaches extremal values ($\chi \to 1$). 
}.
For each run with BH\-aH\-AHA 
output, we compare their measured irreducible mass
to their expected value 
\begin{equation} \label{eqn:m_irr}
  m_{\rm irr} = m_i \sqrt{ \left( 1 + \sqrt{1 - {\chi_i}^2} \right) / 2} 
\end{equation}
(where $m_1 = q/(1+q)$ for the primary and $m_2 = 1/(1+q)$ for the secondary),
and track the relative error as a measure of run quality.

All well-resolved simulations showed good mass conservation up to plunge. 
While BH\-aH\-AHA\ 
was not fully integrated when the $q=24$, $\epsilon_{\psi} = 10^{-3}$ run began, 
Figure~\ref{fig:q24_bah} shows for the $q=24$, $\epsilon_{\psi} = 10^{-4}$ run that at this resolution
the apparent horizon mass remains well-conserved throughout the simulation,
with errors generally below $10^{-4}$ throughout
(consistent with null mass drift to BH\-aH\-AHA's 
measurement precision). 

The well-resolved $q=12$ runs (Fig.~\ref{fig:q12_bah})
at low spin $\chi \leq 0.4$ have similarly low errors, $\lesssim 10^{-4}$. 
At higher spin, the $\chi = 0.8$ runs approach a relative error in the irreducible mass of $\lesssim 10^{-3}$, remaining largely constant despite the small point count across the horizon diameter of the secondary ($N_2 \sim 38$). 
In contrast, the under-resolved, spinless $q=12$ run suffered
significant mass loss ($\sim 2\%$ across the $\sim 1000 \, M$ of the inspiral)
due to insufficient resolution about the larger BH at the start of the simulation (see Table~\ref{tab:runs}).
All other $q=12$ runs had increased resolution about the larger BH by one refinement level, which removed this instability 
(driven by aggressive Hamiltonian damping on the larger BH).

\subsection{Waveform quality}

As we discuss in \S\ref{sec:discuss_WAMR}, 
waveform purity in and of itself is a good
measure of run quality. 
Generally, the lower the noise in the waveform, 
the lower the errors in the simulation. 
In the current work, we quantify waveform noise 
by first subtracting a secular (time-smoothed) fit to the waveform. Second, we measure the amplitude of variations on this signal (e.g. using \href{https://docs.scipy.org/doc/scipy/reference/generated/scipy.signal.hilbert.html}{a Hilbert transform}). If these fluctuations on top of the secular behavior shrink, we consider the waveform noise to have decreased. 
A stark example of this can be seen in 
Figure~\ref{fig:q24_early_cf}.  On better resolving the gauge wave, 
we see a dramatic curtailment in back-reflections,
leading to an order of magnitude reduction 
in waveform noise. 

%

\section{Path reversal} \label{sec:path_reversal}

Due to frame dragging of the larger black hole in retrograde configurations, 
the smaller BH's trajectory is expected to reverse as it approaches merger, 
as shown in \citet{Price+13} for a point particle.

\begin{figure}[H]\centering
  \includegraphics[width=\linewidth]{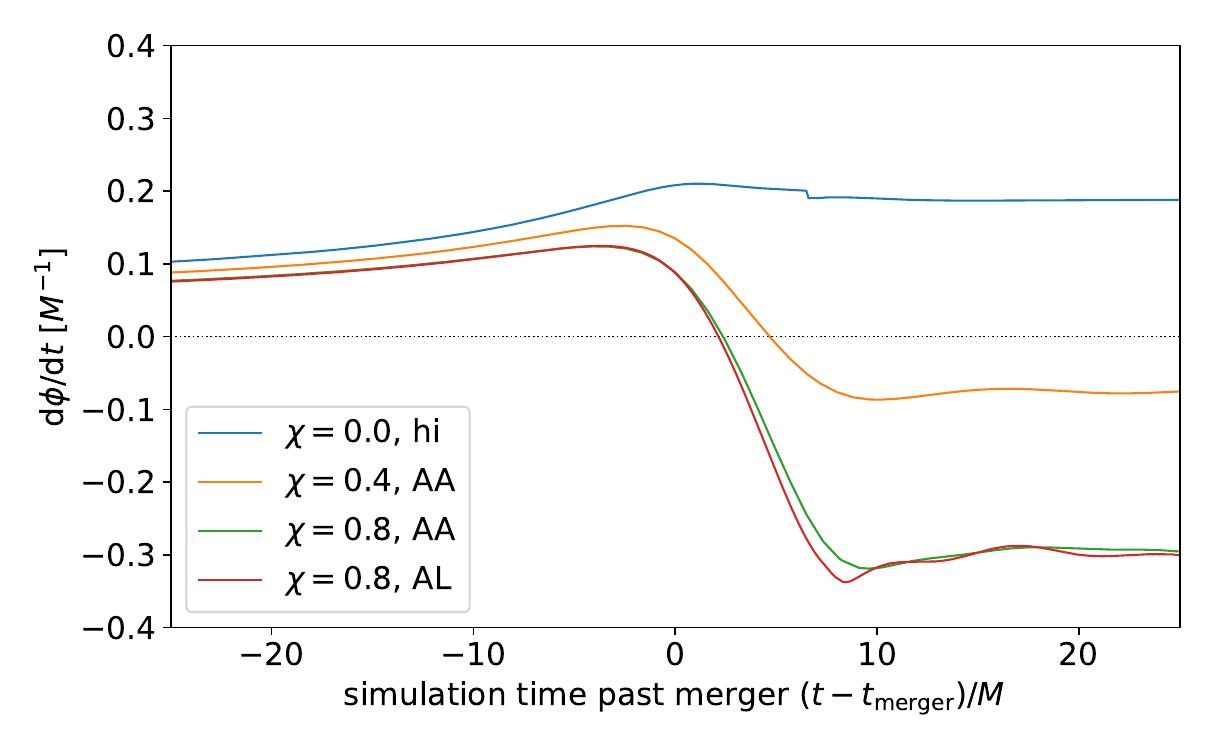}
  \caption{
    Orbital angular velocity for each of the non-precessing $q=12$ runs,
    highlighting the reversal of the inspiral trajectory with our retrograde orbits. 
  }
  \label{fig:path_reversal}
\end{figure}

Figure~\ref{fig:path_reversal} shows the orbital 
angular velocity for the non-precessing $q=12$ runs. 
The zero-crossing for the retrograde configurations 
shows this path reversal in our simulation.

\section{Kick velocities} \label{sec:kick_velocities}

After BHs merge, they can typically have a significant kick---a non-zero velocity of the final, merged BH. 
Spinless configurations range in kick magnitude from roughly $15 \, {\rm km/s}$ near $q=24$ 
to nearly 200 kilometers per second toward more equal masses \citep{Islam+26}. 
However, once large spins become possible, 
post-Newtonian models break down, 
particularly at larger mass ratios with precession. 

Table~\ref{tab:kicks} displays gravitational recoil kick velocity magnitudes for each of the completed runs. 
Most kicks agree with {\sc gwModels} \citep{Islam+26} to $<1 \sigma$. 
The precessing case exceeds predictions by over $2 \sigma$, 
even for the radiated velocity measurement. 
Coordinate velocities are gauge-dependent,
and several are considerably larger than predicted. 
Uncertainties on spinning runs reflect only variation and uncertainty in the fitting process within the individual run. 
We defer further discussion to future work.



\begin{table*}[!hbt]
  \caption{
    Kick velocity (km/s) for all runs.
    The coordinate velocity $v_{\rm coord}$ extrapolates final speeds toward infinite radius ($1/r=0$) and is heavily affected by the gauge. 
    The radiated velocity $v_{\rm rad}$ sums $\Psi_4$ linear-momentum flux. 
    We use {\sc gwModels} to compare against predicted kick magnitudes $v_{\rm pred.}$ (km/s). 
      Nonspinning cases use the analytic fit, 
      (anti-)aligned spin cases use Gaussian Process Regression (GPR) fits, and 
      the precessing case uses the Normalizing Flow (NF; orientation-averaged). 
    The high-res $q=24$ run is marked ``---'', as we did not evolve it through merger. 
  }
  \label{tab:kicks}
  \begin{ruledtabular}
  \begin{tabular}{lccccc}
    Run Label & $q$ & $v_{\rm coord}$ (km/s) & $v_{\rm rad}$ (km/s) & $v_{\rm pred.}$ (km/s) & Fit Type \\
    \hline 
    $\epsilon_{\psi} = 10^{-3}$ & 24 &   $15 \pm 8$ & $15.6 \pm 0.6$ & $15.40 \pm 0.13$ & Analytic \\
    $\epsilon_{\psi} = 10^{-4}$ & 24 & --- & --- & $15.40 \pm 0.13$ & Analytic \\
    $\chi=0.0$, lo & 12 &   $52 \pm 4$ & $48.2 \pm 2.3$ &  $46.9 \pm 0.4$ & Analytic \\
    $\chi=0.0$, hi & 12 & $48 \pm 3$ & $50.2 \pm 1.0$ &  $46.9 \pm 0.4$ & Analytic \\
    $\chi=0.4$, AA & 12 &  $53 \pm 8$ & $57.8 \pm 1.5$ &  $64 \pm 20$ & GPR \\
    $\chi=0.8$, AA & 12 &  $140 \pm 8$ & $66 \pm 3$ & $75 \pm 34$ & GPR \\
    $\chi=0.8$, AL & 12 &  $213 \pm 10$ & $63 \pm 3$ & $77 \pm 23$ & GPR \\
    $\chi=0.8$, PR & 12 & $1680 \pm 100$ & $280 \pm 5$ & $126 \pm 74$ & NF \\
  \end{tabular}
  \end{ruledtabular}
\end{table*}

\bibliographystyle{apsrev4-2}
\bibliography{main.bib}

\end{document}